\documentclass{emulateapj}

\usepackage{natbib}
\begin{document}

\shorttitle{The $z\sim5$ QLF}

\shortauthors{McGreer et al.}
\title{The $z=5$ Quasar Luminosity Function from SDSS Stripe 82
\thanks{Observations reported here were obtained at the MMT Observatory, a joint facility of the Smithsonian Institution and the University of Arizona. This paper also includes data gathered with the 6.5-m Magellan Telescopes located at Las Campanas Observatory, Chile.}}
\author{
Ian D. McGreer\altaffilmark{1},
Linhua Jiang\altaffilmark{2, \dag},
Xiaohui Fan\altaffilmark{1},
Gordon T. Richards\altaffilmark{3},
Michael A. Strauss\altaffilmark{4},
Nicholas P. Ross\altaffilmark{5},
Martin White\altaffilmark{5,6},
Yue Shen\altaffilmark{7},
Donald P. Schneider\altaffilmark{8,9},
Adam~D.~Myers\altaffilmark{10}, 
W. Niel Brandt\altaffilmark{8},
Colin~DeGraf\altaffilmark{11},
Eilat Glikman\altaffilmark{12},
Jian Ge\altaffilmark{13},
Alina Streblyanska\altaffilmark{14,15}
}
\altaffiltext{1}{Steward Observatory, The University of Arizona, 
                 933 North Cherry Avenue, Tucson, AZ 85721--0065, USA}
\email{imcgreer@as.arizona.edu}
\altaffiltext{2}{School of Earth and Space Exploration, Arizona State University, Tempe, AZ 85287, USA}
\altaffiltext{3}{Department of Physics, Drexel University, 3141 Chestnut Street, Philadelphia, PA 19104, USA}
\altaffiltext{4}{Department of Astrophysical Sciences, Princeton University, Princeton, NJ 08544, USA}
\altaffiltext{5}{Lawrence Berkeley National Laboratory, 1 Cyclotron Road, Berkeley, CA 92420, USA}
\altaffiltext{6}{Department of Physics, 366 LeConte Hall, University of California, Berkeley, CA 94720, USA}
\altaffiltext{7}{Harvard-Smithsonian Center for Astrophysics, 60 Garden Street, Cambridge, MA 02138, USA}
\altaffiltext{8}{Department of Astronomy and Astrophysics, The Pennsylvania State University, 525 Davey Laboratory, University Park, PA 16802, USA}
\altaffiltext{9}{Institute for Gravitation and the Cosmos, The Pennsylvania State University, University Park, PA 16802, USA}
\altaffiltext{10}{Department of Physics and Astronomy, University of Wyoming, Laramie, WY 82071, USA}
\altaffiltext{11}{McWilliams Center for Cosmology, Carnegie Mellon University, 5000 Forbes Avenue, Pittsburgh, PA 15213, U.S.A.}
\altaffiltext{12}{Department of Physics and Yale Center for Astronomy and Astrophysics, Yale University, P.O. Box 208121, New Haven, CT 06520-8121}
\altaffiltext{13}{Dept. of Astronomy, University of Florida, 211 Bryant Space Science Center, Gainesville, FL, 32611, USA}
\altaffiltext{14}{Instituto de Astrofisica de Canarias (IAC), E-38200 La Laguna, Tenerife, Spain}
\altaffiltext{15}{Dept. Astrofisica, Universidad de La Laguna (ULL), E-38206  La Laguna, Tenerife, Spain}
\altaffiltext{\dag}{Hubble Fellow}

\begin{abstract}
We present a measurement of the Type I quasar luminosity function at $z=5$ 
using a large sample of spectroscopically confirmed quasars selected from
optical imaging data. We measure the bright end ($M_{1450}<-26$) with 
Sloan Digital Sky Survey (SDSS) data covering $\sim6000~{\rm deg}^2$, then 
extend to lower luminosities ($M_{1450}<-24$) with newly discovered, faint 
$z\sim5$ quasars selected from 235~deg$^2$ of deep, coadded imaging in the 
SDSS Stripe 82 region (the celestial equator in the Southern Galactic Cap). 
The faint sample includes 14 quasars with spectra obtained as ancillary 
science targets in the SDSS-III Baryon Oscillation Spectroscopic Survey 
(BOSS), and 59 quasars observed at the MMT and Magellan telescopes. We 
construct a well-defined sample of $4.7<z<5.1$ quasars that is highly 
complete, with 73 spectroscopic identifications out of 92 candidates. 
Our color selection method is also highly efficient: of the 73 spectra 
obtained, 71 are high redshift quasars. 
These observations reach below the break in the luminosity function 
($M_{1450}^*\approx-27$). The bright end slope is steep ($\beta \la -4$),
with a constraint of $\beta < -3.1$ at 95\% confidence. The break
luminosity appears to evolve strongly at high redshift, providing an
explanation for the flattening of the bright end slope reported
previously.
We find a factor of $\sim2$ greater decrease in the number 
density of luminous quasars ($M_{1450}<-26$) from $z=5$ to $z=6$ than from 
$z=4$ to $z=5$, suggesting a more rapid decline in quasar activity at high 
redshift  than found in previous surveys. Our model for the quasar 
luminosity function predicts that quasars generate $\sim30$\% of the 
ionizing photons required to keep hydrogen in the universe ionized at $z=5$.
\end{abstract}


\clearpage

\section{Introduction}\label{sec:intro}

The number density of quasars evolves strongly with redshift,
a conclusion reached shortly after the initial identification of
cosmological redshifts for quasars. 
Quasars increase in number with increasing redshift \citep{Schmidt68}
until $z\sim2.5$, when quasar activity peaks,
as surveys for higher redshift quasars show a steep decline
in number \citep{Osmer82,WHO94,SSG95,Fan+01LF,Richards+06LF}. 
Quasars are associated
with accretion onto supermassive black holes \citep{Salpeter64};
how the first such black holes grew from initial seeds and were
triggered by gas accretion to become luminous quasars remain key
questions about the evolution of the early universe
\citep[see, e.g., the recent review by][]{Volonteri10}. Indeed,
quasars are now observed to $z\sim7$ \citep{Mortlock+12},
indicating that the mechanisms that drive them were in place
within 0.8 Gyr after the Big Bang.

The quasar luminosity function (QLF) is one of the most fundamental
observational probes of the growth of supermassive black holes over
cosmic time. The QLF is generally found to have the form of a
broken power law \citep{BSP88,Pei95,Boyle+00}, 
with a steep slope towards high luminosities
and a flatter slope extending to low luminosities. 
Observed evolution in the shape of the QLF with redshift --- e.g.,
a change in the power law slopes or the location of the break
luminosity --- may provide insight into the physics of black hole
growth. Many studies have found
evidence for evolution of the QLF in large quasar samples 
\citep{SSG95,Fan+01LF,Richards+06LF,HRH07,Croom+09}. 
One key consequence of this evolution is the observed ``downsizing'' of 
quasar activity: where by the spatial density of more luminous 
objects peaks at higher redshifts 
\citep{Cowie+03,Ueda+03,Hasinger+05,Croom+09}.

Theoretical models for quasars have explored a variety of physical
processes that may drive the triggering of quasar activity. 
These models make testable predictions about the evolution
of the QLF based on the evolution of the triggering mechanisms.
For example, models where quasar activity is instigated by
mergers of gas-rich galaxies 
\citep[e.g.,][]{Hernquist89,Carlberg90,CHR99,KH00,Hopkins+06unified,Shen09}
tie the QLF to the evolution of the merger rate of dark matter
halos in cosmological simulations. Alternatively, some 
hydrodynamical simulations point to rapid inflows of cold gas along 
filamentary structures in the cosmic web as the primary fueling 
mechanism at high redshift \citep[e.g.,][]{DiMatteo+12}. 
It is also possible to predict the QLF while being agnostic to the 
specific triggering mechanisms; \citet{CW12} describe a 
model for populating galaxies with accreting black holes that
directly relates the evolution of quasars to that of their host galaxies.

Furthermore, feedback from black hole accretion is expected to play 
an important role in regulating the growth of black holes and the 
duration of quasar activity \citep[e.g.,][]{DSH05,Hopkins+06unified}. 
As the QLF is a convolution of the black hole mass function and the
Eddington ratio\footnote{The ratio of the mass accretion rate onto the
black hole to the maximal rate allowed by the Eddington limit.} 
distribution, these processes will
alter the shape of the QLF \citep{Hopkins+05lifetimes}.
While the low-redshift QLF is reasonably well measured by surveys
at optical, X-ray, and mid-infrared wavelengths 
\citep[see, e.g., the compilation of][]{HRH07}, observations at
high redshift are less constraining on these models, even
though it is at high redshift where, for example, different feedback
models make significantly different predictions for the quasar
population \citep{Hopkins+07clustering}. Feedback-regulated models
for quasar activity tend to predict strong evolution of the
bright-end slope with redshift \citep[e.g.,][]{KH00,WL03}, and also
provide a physical explanation for the observed downsizing trend 
\cite{SSB05}.

Measurements of the high redshift QLF also allow estimates of
the contribution of quasars to the ionization state of the 
intergalactic medium (IGM) during and after reionization. Quasars 
are unlikely to produce enough ionizing photons to be the primary driver 
of reionization \citep{Fan+01PI} although current constraints on the 
quasar ionizing photon budget are limited to extrapolations from the 
bright end of the QLF and a handful of faint $z\sim6$ quasars 
\citep{Jiang+09,Willott+10}, as well as upper limits on moderate
luminosity AGN from X-ray surveys \citep{Barger+03,Fontanot+07}. 
Quasars do have a
much harder spectrum than stellar sources, but constraints from
the soft X-ray background limit the contribution of high energy
photons to large-scale reionization \citep{DHL04,McQuinn12}. Nonetheless,
there are few observational constraints on the faint quasar population during
the epoch of reionization. And while quasars may not be directly 
responsible for hydrogen reionization, they 
are expected to provide the
high-energy photons responsible for \ion{He}{2} reionization at lower
redshifts \cite[e.g.,][]{HL98,MHR99,MHR00}.
Models of IGM evolution thus benefit from improved
observational constraints on quasar activity at high redshift
\citep[e.g.,][]{BOF09}.

Currently, only optical and near-IR imaging surveys provide the
requisite area and depth to construct large samples of quasars at high 
redshifts \cite[see, e.g., the compilation given in][]{Ross+12qlf}.
Throughout most of the 1990s, the redshift record was held by a 19th
magnitude quasar at $z=4.9$ \citep{SSG91}. The first quasar with $z\ge5$
was discovered by \citet{Fan+99} in commissioning data from the 
Sloan Digital Sky Survey \citep[SDSS;][]{York+00}. The same data
were used to build a sample of 39
quasars with $3.6<z<5.0$ to a limit of $i=20$ \citep{Fan+01LF} and to
study the evolution of quasars at high redshift, confirming the steep
decline in number density at $z>3$ (roughly a factor of three per unit
redshift). The early high-redshift quasar surveys also found evidence
for flattening of the bright-end slope relative to lower redshifts
\citep{KK88,SSG95,Fan+01LF}, a result seemingly confirmed by a large,
homogeneous quasar sample from the SDSS extending to $z=5$ 
\citep{Richards+06LF}. This form of evolutionary trend in the 
bright-end slope would contribute to a downsizing effect, by slowing
the decline in number density with redshift for quasars above the 
break luminosity. 

The SDSS-III Baryon Oscillation Spectroscopic Survey 
\citep[BOSS;][]{Eisenstein+11,Dawson+12} aims to collect spectra
of over 150,000 quasars with redshifts between 2.2 and 3.5. 
\citet{Ross+12qlf} present
the QLF measured from 22,000 color-selected quasars from BOSS DR9, 
as well as two samples of variability-selected quasars at $z<3.5$
from \citet{PD11} and \citet{PD12}.
The BOSS spectroscopic target selection extends $\sim1.5$ mag
fainter than the SDSS, 
reaching below the break in the luminosity function to $z\sim3.5$. 
Combined with the 2SLAQ \citep{Croom+09}, these large quasar surveys
form a fairly complete picture of the optically unobscured (Type I) quasar
population at $z\la3$.

At higher redshifts, over 50 quasars are now known at $z\ga6$.
While constraints at these redshifts are weaker (and await future
wide-area surveys for greater numbers), the best determinations
to date indicate that the bright-end slope is steep 
\citep{Jiang+08,Willott+10}, roughly agreeing with the $z<3.5$
measurements from BOSS and the $z\sim2.5$ determination from
\citet{Croom+09}. One of the aims of this work is to
examine the QLF at intermediate redshifts and test previous
claims for a flattening of the bright-end slope at $z>3$.

We present a measurement of the QLF at $z=5$, combining
bright quasars from the SDSS
with faint quasars reaching nearly
2 mag deeper. These quasars are drawn from optical imaging
data (probing the rest-frame ultraviolet) and thus have low 
intrinsic extinction, and are confirmed
by spectroscopy to be broad emission line, Type I quasars.
The faint sample is derived from coadded optical 
imaging in the SDSS Stripe 82 region. This imaging covers
$235~{\rm deg}^2$ to a depth more than 2 mag fainter than the
SDSS main survey. The combination of medium depth and medium
sky area -- relative to SDSS and to small-area deep fields --
is ideal for searching for rare, faint high-redshift quasars.
With Stripe 82 we are able to reach sufficient depth to probe the
faint end of the luminosity function, while attaining enough dynamic 
range to constrain its overall shape. This uniform sample of $z\sim5$ 
quasars --- selected with simple criteria over a homogeneous
imaging area --- is larger than all $z\ga6$ surveys combined,
and provides a key link in our understanding of quasar evolution
at high redshift.

\begin{figure*}[!t]
 \epsscale{1.15}
 \plotone{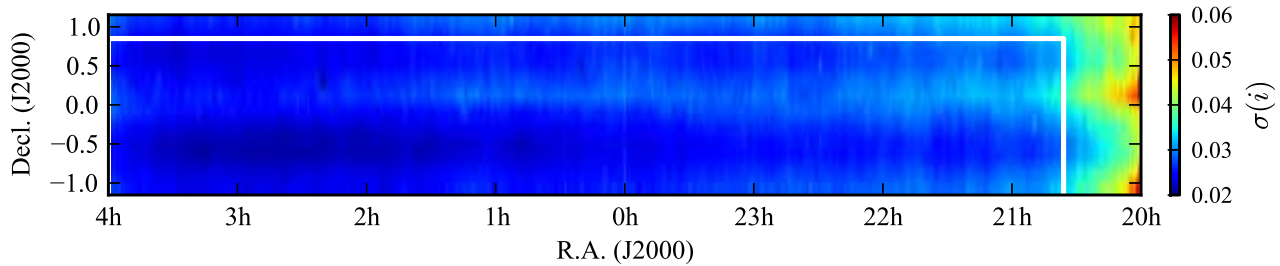}
 \caption{
  Median photometric uncertainty for stars with $i=22$ along
  Stripe 82. The depth is very uniform, and with a typical uncertainty
  of $\sigma(i) \approx 0.03$ at the flux limit for our quasar survey the
  completeness is high. We search for quasar candidates within the
  region bounded by the white lines. The cutoff at 
  R.A. = $20^{\rm h}~32^{\rm m}$ is due to the high stellar density
  along the western edge of Stripe 82, as well as the lack of UKIDSS
  coverage there. Note that the photometric uncertainties increase outside
  the boundary, likely due to crowding in the high stellar density regions.
  The cutoff at $\delta=0.85^\circ$ is due to problems in
  the $g$-band coadded images in the northernmost two camera columns
  of the Stripe.
 \label{fig:stripe82map}
 }
\end{figure*}

We first describe the coadded optical images on Stripe 82 that
provide the primary catalogs from which we select candidates, as well
as infrared imaging from UKIDSS and our own observations used
to reduce stellar contamination (\S~\ref{sec:imaging_data}). 
Section~\ref{sec:candidate_selection} outlines our selection criteria, 
derived from models of quasar colors, and based on simple color cuts. 
We combine spectroscopy
from the SDSS and BOSS surveys with our own
observations using the MMT and Magellan telescopes to build a
sample of over 70 $z\sim5$ quasars to a limit of $i_{\rm AB}=22$;
the observations and data processing are presented in section~\ref{sec:specobs}
and the catalog of quasars is provided in section~\ref{sec:catalog}. In section~\ref{sec:results}
we use our quasar color model to quantify the completeness of the
survey; with this in hand we calculate the binned luminosity
function and derive a parametric form for the luminosity function
using the maximum likelihood technique. We explore the evolution of
the QLF at high redshift and determine the contribution of quasars
to the ionizing background at $z\sim5$ based on our QLF model.
We present conclusions
in section~\ref{sec:conclusions}.

We incorporate photometric data from several sources in this work; for 
consistency, all magnitudes are converted to the AB system 
\citep{OkeGunn83} unless otherwise noted. SDSS reports $ugriz$ photometry 
on the asinh scale \citep{Lupton+99}, which is nearly identical to AB
at bright magnitudes. Although we use SDSS photometry only for bright 
quasars, we will note any instances where the asinh system is used. We refer 
to the SDSS DR7 imaging as ``SDSS main'', in contrast to the deep imaging
we use to build our primary faint quasar sample, referred to as 
``Stripe 82'' or ``the coadd imaging'' (\S\ref{sec:coadd_imaging}).
UKIDSS magnitudes are Vega-based, but we convert all
UKIDSS magnitudes to AB using the values given in \citet{Hewett+06}. 
All magnitudes are corrected for
Galactic extinction \citep{SFD98} unless otherwise noted.
We use a $\Lambda$CDM cosmology with parameters
$\Omega_\Lambda=0.728$,~$\Omega_m=0.272$,~
$\Omega_{\rm b}=0.0456$,~and~
$H_0=70~{\rm km}~{\rm s}^{-1}~{\rm Mpc}^{-1}$
\citep{Komatsu+09}.

\begin{figure}[!ht]
 \epsscale{1.1}
 \plotone{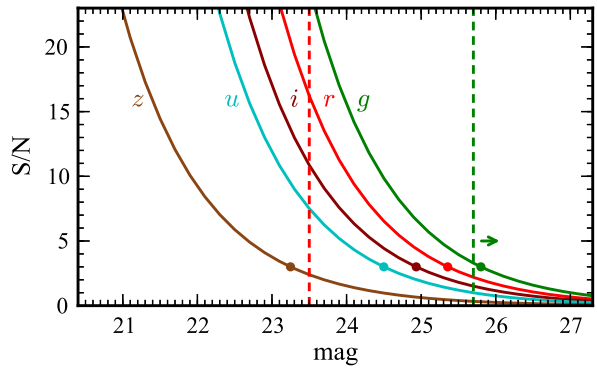}
 \caption{
  Stripe 82 coadded imaging depth in the five SDSS bands. 
  Lines represent
  the median $S/N$ measured in 1.8\arcsec\ apertures for isolated stellar
  objects (no extinction correction has been applied).
  Filled circles are drawn at the $3\sigma$ limits.
  The vertical dashed lines show the approximate $r$ (red) and
  $g$ (green) magnitudes for a typical $z\sim5$ quasar at our survey limit
  of $i=22$, based on colors of $r-i=1.5$ and $g-r \ge 2.2$. By comparison,
  the 95\% completeness limits for single-epoch SDSS imaging are
  $g\approx22.2$, $r\approx22.2$, and $i\approx21.3$~\citep{Stoughton+02}.
  The depth of the coadded imaging in the $g$ band allows for reliable
  selection of very red objects.
 \label{fig:coadd_depth}
 }
\end{figure}

\section{Imaging Data}\label{sec:imaging_data}

\subsection{Stripe 82 coadded imaging}\label{sec:coadd_imaging}

Over the ten year duration of the survey, the SDSS I/II repeatedly imaged
a 2{\fdg}5$~\times~100^\circ$ stripe centered at zero declination in the 
Southern Galactic Cap \citep{dr7}. This repeat imaging increased in frequency
during the latter part of operations as part of the SDSS Supernova 
Survey \citep{Frieman+07}. By the end of operations, over 100 repeat
imaging scans were obtained along Stripe 82 
\citep[its designation according to the survey geometry,][]{Stoughton+02}.
As with the main survey, the imaging was obtained with a
drift-scan camera \citep{Gunn+98} mounted on the 2.5m Sloan telescope \citep{Gunn+06},
with nearly simultaneous 54.1 s exposures in five broad optical 
bands \citep[$ugriz$;][]{Fukugita+96}. The quality restrictions
applied to the main survey imaging
\citep[photometric and good-seeing conditions;][]{Ivezic+04,Hogg+01}
were relaxed for the Supernova Survey in order to increase the
temporal coverage, resulting in a greater range of data quality.

\citet{Jiang+09} describe deep $riz$ images of Stripe 82 constructed
from the coaddition of 50-60 individual imaging scans at each position
on the sky. Images
with seeing poorer than 2\arcsec\ or an $r$-band sky background
brighter than 19.5~mag~arcsec$^{-2}$ were rejected. The remaining
images were weighted by their transparency, seeing, and sky
background noise, and coadded using the SWARP software \citep{swarp}.
The final coadded images reach a depth roughly 2 mag fainter
than single-epoch SDSS images, and were used to discover $z\sim6$ quasars 
as faint as $z_{\rm AB}=22.2$ \citep{Jiang+09}. These data are
not the only coadded images available for Stripe 82: \citet{Annis+11}
produced coadded images made using a similar process but with stricter 
image quality criteria resulting fewer imaging scans used at each sky
position, while \citet{Huff+11} created image
stacks optimized for weak lensing shear studies. 

Our starting point is the $riz$ coadded images from \citet{Jiang+09},
which we now extend to include the $u$ and $g$ bands
(Jiang et al., in preparation). The deep
imaging in the bluer SDSS bands is being used to discover
ultra-luminous Lyman break galaxies at $z>2.5$ (Bian et al., in preparation).
In this work, we extend the \citet{Jiang+09} search for faint, $z\sim6$
quasars to lower redshifts ($z\sim5$),
as the deep $g$-band photometry combined with the redder bands allows
reliable color selection of quasars in this redshift range.
The depth of our coadds is similar to that of the \citet{Annis+11} coadds
(see their Fig. 7); however, as described in \S\ref{sec:boss_spec} we
encountered much greater contamination when using the catalogs from
the \citet{Annis+11} coadds for quasar selection.

We use Sextractor \citep{sextractor} to produce object catalogs from the
deep $ugriz$ images. Object detection was performed in the
$i$-band, and fluxes and other measurement parameters in the other 
bands were derived from apertures centered on the $i$-band object
position using the dual-image mode of Sextractor. All fluxes
and magnitudes from the coadded Stripe 82 imaging quoted in 
this work are taken from aperture photometry using a 1.8\arcsec\ diameter. 
The typical seeing in the coadded images\footnote{The weighting scheme 
used when coadding the images favors better seeing data; it is
$w=T\times({\rm FWHM}^2\sigma^2)^{-1}$, where $T$ is the transparency, FWHM
is the full width at half-maximum of the point-spread function (PSF),
and $\sigma^2$ is the variance of the sky background \citep{Jiang+09}.} 
is 1.4\arcsec, 1.3\arcsec, 1.1\arcsec, 1.0\arcsec,
and 1.1\arcsec\ in $u$, $g$, $r$, $i$, and $z$, respectively.

The geometry of the Stripe 82 coadded imaging follows that of the
parent SDSS imaging\footnote{The SDSS camera consists of six columns
of CCDs that are continuously exposed in a drift-scan mode \citep{Gunn+98}. 
The gaps between the CCD columns are filled by combining two Strips
offset in declination (each consisting of a single scan) to form a Stripe. 
Thus the declination axis of Stripe 82 is split into 12 columns during the 
imaging scans. Each scan is 13\arcmin\ wide in declination, broken into
10\arcmin\ segments along right ascension; these image segments are 
referred to as fields.}.
We photometrically calibrated individual fields in the coadded catalogs
by deriving zero points from the \"ubercalibrated \citep{Padmanabhan+08} 
SDSS DR8 \citep{dr8} photometry (derived from the best individual 
imaging scan at each location in Stripe 82), using stars with 
$15.5 < r < 20.5$. We thus tie the aperture photometry from Sextractor
on a field-by-field basis to PSF photometry from SDSS. This process results
in zero points that account for seeing and photometric variations between 
the fields, but not within the fields. However, the large number of images 
contributing to the coadds tends to average over effects such as 
varying sky backgrounds and PSF shapes. In general, we find the 
photometry is highly consistent between fields and agrees quite well 
with SDSS photometry for the brighter objects. 
Figure~\ref{fig:stripe82map} shows that the depth across the Stripe is 
highly uniform, and Figure~\ref{fig:coadd_depth} shows the depths
reached in the $ugriz$ bands in the coadded imaging.

Quasars at $z\sim5$ have similar fluxes in the $i$ and $z$ bands.
We adopt the $i$-band as our detection band as it provides greater
a greater signal-to-noise ratio. However, at $z\sim5$ the Ly$\alpha$ 
emission line and the Ly$\alpha$ forest are within the $i$-band. The 
fluxes in the band that defines our detection threshold are thus subject 
to the substantial variances in Ly$\alpha$ equivalent widths,
the incidence of strong Ly$\alpha$ absorption systems, and 
variations in the mean IGM opacity. The $z$-band is less affected 
by these issues (although it does contain \ion{C}{4} in emission); 
however, it is  significantly noisier than the $i$-band because of 
lower CCD sensitivity and a higher sky background. We will discuss
these issues further in \S\ref{sec:catalog}.

\begin{figure*}[!t]
 \epsscale{1.1}
 \plotone{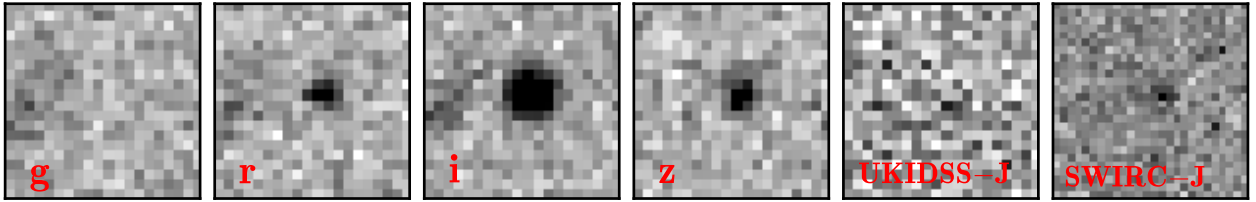}
 \caption{Example images of a typical faint quasar at $z=4.9$ selected by this 
  survey. The panels contain the $griz$ coadded imaging, followed by the UKIDSS
  $J$ image, and a $J$ image from MMT/SWIRC. This object (J234601.5-003855)
  is not detected in the $g$-band, and has $r=23.64\pm0.09$, $i=21.72\pm0.02$, 
  and $z=21.74\pm0.08$.
  It is undetected in UKIDSS; the SWIRC imaging yields $J=21.39\pm0.13$.
  In the best single-epoch imaging from SDSS DR8, this object is a $\sim9\sigma$
  detection in the $i$-band and $\la3\sigma$ in $r$ and $z$.
 \label{fig:cutout}
 }
\end{figure*}

We further note that our survey is restricted to less than the full area 
on Stripe 82 for two reasons. At the time that we performed
target selection, the coadds corresponding to the two uppermost 
camera columns had problems with the sky background in the
$g$-band, rendering them inadequate for our purposes. We thus
imposed a cutoff of $\delta < 0.85^\circ$ when selecting candidates.
In addition, the western edge of Stripe 82 (near 20$^{\rm h}$)
approaches the Galactic plane. Attempting color selection of
quasars in this region would result in overwhelming
stellar contamination \cite[see, e.g., Figure 1 of][]{VandenBerk+05}. We
thus restrict our survey to objects with R.A. $>~20^{\rm h}~32^{\rm m}$
(this region also lies outside the UKIDSS imaging area described in
the next section).
The final area of the deep imaging we used
for $z\sim5$ quasar selection is $235~{\rm deg}^2$, extending from
$20^{\rm h}~32^{\rm m}$ to $4^{\rm h}$, and from $-1.25^\circ$ to
$+0.85^\circ$~(Fig.~\ref{fig:stripe82map}).

\subsection{UKIDSS}\label{sec:UKIDSS_imaging}

The UKIRT Infrared Deep Sky Survey \citep[UKIDSS; ][]{Lawrence+07} 
consists of multiple
infrared imaging surveys with the UKIRT Wide Field Camera \citep[WFCAM; ][]{Casali+07}, including a wide-area component 
(the Large Area Survey or LAS) that covers 
$\sim4000~{\rm deg}^2$ to a depth of  
$J_{\rm AB} \approx 20.6$ ($5\sigma$).
The LAS includes Stripe 82, providing shallow infrared imaging of
our candidates and additional leverage in discriminating high redshift
quasars from stellar contaminants.
We first apply a loose color cut to identify potential quasar
candidates from the coadded optical imaging, 
\[
  (r-i) > 1.0~~\&\&~~(i-z) < 0.625[(r-i)-1.0] + 0.2~,
\]
then queried the DR8plus 
release in the WFCAM Science Archive \citep[WSA; ][]{Hambly+08} for infrared counterparts.

Quasars at $z\sim5$ have $(i-J)_{\rm AB} \sim 0$ 
(see section~\ref{sec:candidate_selection}), thus at the faint 
limit of our survey most of our target objects do not have a
counterpart in the UKIDSS catalogs. However, stars with similar
optical colors to $z\sim5$ quasars (namely, M and L dwarf stars)
are redder in the near-IR ($0.5 < (i-J)_{\rm AB} < 1.5$, see 
Fig.~\ref{fig:color_color}) and thus are relatively brighter at infrared 
wavelengths. We downloaded $J$-band image cutouts from the WSA for 
all of the candidates pre-selected from the optical imaging,
then performed aperture photometry using the IRAF 
\textit{aper} task at the $i$-band 
position from the SDSS coadded imaging, measured in 2\arcsec\ diameter
apertures. We validated our photometry by checking against the
UKIDSS catalog measurements for brighter objects (utilizing the
UKIDSS calibration as described in \citealt{Hodgkin+09}). 
The aperture photometry reaches $J \approx 21.3$ at $3\sigma$,
sufficient to discriminate the typical stellar contaminants at our
limit of $i=22$.

\subsection{MMT SWIRC}\label{sec_mmtswirc}

At the time the initial candidate selection was performed,
the publicly available UKIDSS data on Stripe 82 did not extend to
R.A. $<21^{\rm h}~36^{\rm m}$. We imaged some of the
candidates lacking $J$ coverage on 2011 Oct 14-15 with the
MMT Smithsonian Widefield Infrared Camera \citep[SWIRC;][]{Brown+08}. 
Each object was observed in the $J$ band using a 9-point dither 
pattern with 30s exposures at each position, for a total integration 
time of 4.5 min.
The typical seeing was 0.6-0.9\arcsec. Images
were dark-corrected, flat-fielded, sky-subtracted, shifted, and combined
using standard IRAF routines. Object photometry was measured
in 2\arcsec\ diameter apertures using IRAF routines, and calibrated
using observations of UKIRT Faint Standards.

A total of 38 objects were observed with SWIRC.
This includes 10 objects already spectroscopically confirmed as
quasars but that lacked UKIDSS coverage or had low $S/N$ in the
$J$-band. These
objects were observed as a check on our $J$-band color selection 
for quasars selected solely by optical colors. The remaining
SWIRC targets were taken from a sample of objects selected as
high-redshift quasar targets based on optical colors but that
lacked UKIDSS coverage. These objects are mainly at 
$309^\circ < \alpha_{\rm J2000} < 360^\circ$, the region of our 
survey with the highest stellar density. Nearly all of these sources 
were readily identified as stellar contaminants from bright detections
in the SWIRC imaging, and would have greatly reduced the
purity of our spectroscopic sample had they not been observed
with SWIRC.

Figure~\ref{fig:cutout} displays postage stamp images of a typical
quasar within our survey. The $i$-band clearly has the highest
$S/N$. The non-detection in UKIDSS and the faint detection in the
deeper SWIRC imaging are expected for a $z\sim5$ quasar; a
contaminating star would have been easily identified from the
infrared imaging.

\section{Quasar Candidate Selection}\label{sec:candidate_selection}

\begin{figure*}[!t]
 \epsscale{1.15}
 \plotone{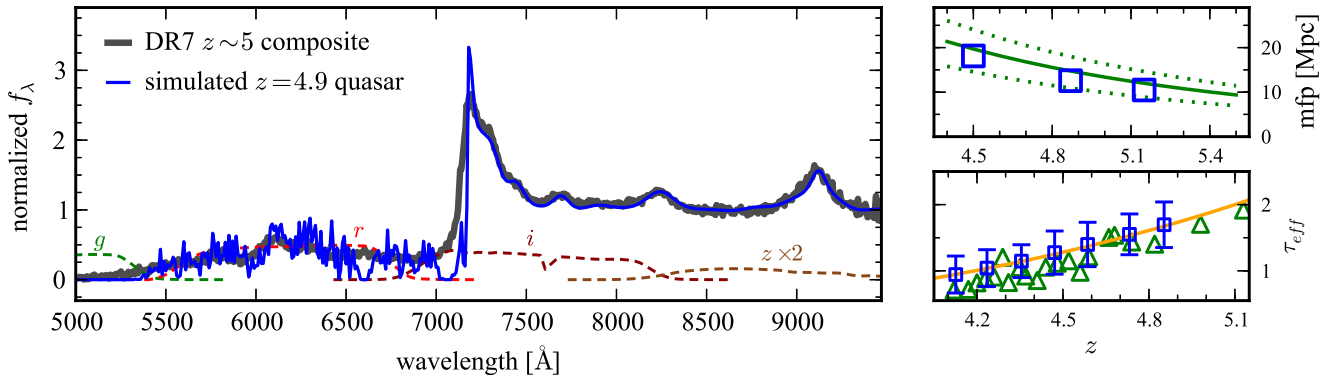}
 \caption{Left panel: example simulated quasar spectrum at $z=4.9$ (blue) 
  compared to a composite spectrum of DR7 quasars with $4.7<z<5.1$ (gray,
  shifted to $z=4.9$).
  The simulated quasar spectrum was randomly drawn from the simulation
  grid described in \S\ref{sec:simulations}, and selected to have 
  $M_{1450}=-26.5$, roughly the median luminosity of $z\sim5$ DR7 quasars. 
  The simulated Ly$\alpha$ forest is calculated at a resolution of $0.1$~\AA; 
  here the spectrum is rebinned to $10$~\AA. This line of sight includes two 
  strong DLAs ($\log{N}_{\rm HI}\sim21.1$) at $z=4.428$ and $z=4.808$.
  The SDSS filter curves are shown as a reference for which parts of the
  spectrum contribute to the broadband colors.
  Upper right panel: the mean free path in proper Mpc from the simulations,
  represented by blue squares. The solid green line represents the evolution in
  the mfp derived by \citet{SC10}, along with the $\pm1\sigma$ range
  (dotted lines).
  Lower right panel: the effective optical depth ($\tau_{\rm eff}$) from the
  simulations, represented as blue squares with error ranges indicating the
  scatter among 1000 simulated lines-of-sight. 
  Measurements from \citet{Songaila04} are 
  shown as green triangles, and the evolution derived by \citet{Fan+06} is shown
  by a solid orange line. We converged on a forest model with mfps $\sim10\%$
  lower than the mean values from \citet{SC10}; however, the $\tau_{\rm eff}$ 
  measurements are highly consistent with the observations.
 \label{fig:simspec}
 }
\end{figure*}

Quasars are typically selected from imaging data using their colors,
but these colors are a strong function of redshift, and 
overlap with the stellar locus at certain redshifts
\citep{Fan99,Richards+01colors,Richards+02sel}.
Optimal techniques have been developed for extracting the much rarer 
quasar population from the overwhelming contamination due to stars 
within the Galaxy. The probability that an object is a quasar or star can
be determined by considering the local density of quasars and stars
in the multidimensional space defined by the set of fluxes available from
an imaging survey. These probabilities can be used to efficiently select
quasar candidates \citep{Bovy+11,Kirkpatrick+11}. 
Similar techniques have also been
applied at high redshift, where the problem is worsened by the
extreme rarity of quasars and the small number of bands in which
the candidates are detected, typically at low $S/N$ \citep{Mortlock+12}. 

Fortunately, at $z\sim5$ quasar colors are well separated from
those of stars. At $z>4.6$ the $r$-band is completely within the 
Ly$\alpha$ forest and the $r-i$ colors of quasars shift markedly 
redward of the stellar locus,
while the $i-z$ colors sample the Ly$\alpha$ to \ion{C}{4} region
of the rest-frame UV and are bluer than the stellar locus.
The $g$-band samples the Lyman limit; $g-r$ colors are generally
redder than those of stars due to the mean forest absorption; 
furthermore, most quasars will encounter a Lyman Limit System (LLS) 
near the red edge of the $g$-band,
leading to saturated absorption and a $g$-band non-detection.
Simple color selection at these redshifts was used in the SDSS I/II
\citep{Richards+02sel},
leading to the first quasars discovered at $z>5$ \citep{Fan+99} and 
a total of over 300 $z>4.6$ quasars by the DR7 release \citep{dr7qso}.

In this section, we describe a model for quasar colors based on
simulated quasar spectra. We use these model quasar colors
to motivate our simple color cuts that achieve a high degree
of both completeness and purity.

\subsection{Simulated quasars}\label{sec:simulations}

\citet{Fan99} outlines a procedure for generating simulated
quasar spectra using a simple empirical model for their
spectral properties at UV/optical wavelengths.
The simulated spectra are integrated through survey
bandpasses to generate simulated fluxes (colors) that 
can be used to define selection criteria, 
and to estimate their completeness. This requires that the spectral
model reliably captures the diversity of quasar spectral features, and
thus accurately reproduces observed quasar colors. 
The basic components of the quasar model are a broken 
power-law continuum, prominent UV/optical emission lines, pseudo-continuum
from Fe complexes, and redshift-dependent Ly$\alpha$ forest absorption 
due to intervening neutral hydrogen. We have used an updated version
of this model (described below) to
reproduce the colors of $\sim60,000$ quasars from the SDSS-III/BOSS
in the range $2.2 < z < 3.5$ \citep{Ross+12qlf}.
We apply the model at higher redshift under the assumption
that the distribution of quasar SEDs does not evolve with redshift
\citep{Kuhn+01,Yip+04,Jiang+06}.
This approach can
be compared to, e.g., cloning the spectra of lower redshift
quasars to higher redshift \citep{Chiu+05,Willott+05}, which carries with
it any selection function imprinted on the lower redshift sample; or
using a small set of quasar templates \citep{Mortlock+12}, which may
not capture objects with unusual features. 

Each quasar is assigned a power law continuum with a break at 1100 \AA. 
The blue slope is drawn from a normal distribution with $\mu(\alpha) = -1.7$ 
and  $\sigma(\alpha) = 0.3$ \citep{Telfer+02}\footnote{Using a larger sample
of quasars with {\it HST Cosmic Origins Spectrograph} observations,
\citet{Shull+12uv} derived a far-UV slope of $\alpha_{\nu}=-1.4$ and a
break wavelength at 1000~\AA. We kept the softer slope from 
\citet{Telfer+02} for consistency with the BOSS analysis. At $z=5$ the 
far-UV colors are dominated by Ly$\alpha$ forest absorption; thus the 
slight change in slope will have little effect on quasar selection.};
the distribution for the red slope is
$\mu(\alpha) = -0.5$ and $\sigma(\alpha) = 0.3$. We add to this continuum
emission lines with Gaussian profiles, where the Gaussian parameters
(wavelength, equivalent width, and FWHM) are drawn from normal
distributions. These distributions are derived from fitting composite
spectra of BOSS quasars in luminosity bins. These distributions 
recover trends in the mean and scatter of the line parameters as a function 
of continuum
luminosity, e.g., the Baldwin Effect \citep{BEff}, and blueshifted lines 
\citep{Gaskell82,Richards+11}.
Finally, we include Fe emission using the template of \citet{VW01}, scaling
the template in segments to match the Fe emission in the composite spectra.
We do not include a contribution from quasars with unusually weak
emission lines, which account for $\sim6$\% of quasars at $z>4.2$ 
\citep{DiamondStanic+09}, or from Broad Absorption Line (BAL) quasars,
which tend to have moderately redder colors 
\citep[e.g.,][]{Weymann+91,Brotherton+01,Reichard+03}.

\begin{figure*}[!t]
 \epsscale{1.15}
 \plotone{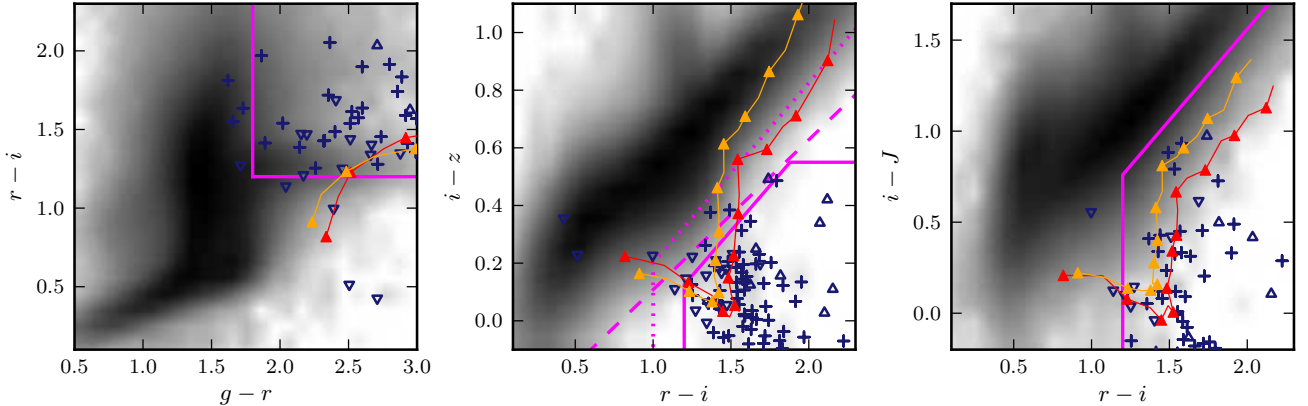}
 \caption{
 Selection criteria for $z\sim5$ quasars. The background image
 shows the density of isolated point-like objects on Stripe 82 
 with $i<22$ and $S/N(u)<2.5$ (i.e., $u$-band dropouts); a
 logarithmic scaling has been applied. These objects
 represent the red end of the stellar locus and are our primary
 source of contamination. Overlaid are the mean color tracks for
 quasars from $z=4.5$ to $z=5.5$, in steps of $\Delta z=0.1$.
 The colors are derived from our simulations and include the
 luminosity dependence of emission line strengths; the two
 tracks are for $M_{1450}=-26.7$ (orange) and $M_{1450}=-24.5$ (red).
 The solid magenta lines are the color selection criteria used
 to identify $4.7 < z < 5.1$ quasars on Stripe 82. In the middle
 panel ($riz$ colors), the dashed line shows the SDSS criteria used
 to target $z>4.6$ quasars and adopted here for the analysis of 
 DR7 quasars, and the dotted line shows our pre-selection criteria
 used to reduce the list of point-like objects for which we obtain
 aperture photometry in the UKIDSS $J$ band. 
 Finally, the purple points show spectroscopically confirmed $z>4.5$ 
 quasars on Stripe 82, with downward pointing triangles denoting
 $z<4.7$ quasars, upward pointing triangles $z>5.1$ quasars, and
 crosses quasars within our primary redshift range of $4.7<z<5.1$.
 \label{fig:color_color}
 }
\end{figure*}

The quasar model for the BOSS employs a prescription for the Ly$\alpha$ 
forest based on the work of \citet{WP10}. This forest model is calibrated to 
observations at $z<4.6$. We extend the model to higher redshifts by using the 
observed number densities of high column density systems from \citet{SC10}.
For the lower column density systems ($\log N_{HI} < 17.2$), we begin with the
parameters of \citet{WP10}, and follow their method of deriving the number 
densities by simulating a large number of sight lines and matching the mean 
free paths (mfps) of the simulated sightlines to the observations
of \citet{SC10}. We also force continuity between the column density
distribution function ($dn/dN_{\rm HI}dz$) at low and high column densities, 
and check the derived effective optical depth ($\tau_{\rm eff}$) against the 
measurements of \citet{Songaila04} and \citet{Fan+06}.
There remains significant uncertainty in forest parameters at high redshift,
but our simple color selection method is relatively insensitive to
the model for the Ly$\alpha$ forest. To account for the degree of uncertainty
in the forest absorption, we also include a ``low'' and ``high'' forest model,
scaling the number densities of forest absorbers up and down by 10\%,
matching the scatter in measurements by \citet{SC10}.
After comparison with observed quasar colors, we find that the model with
a mean forest density 10\% greater than the best-fit value from the mfp
measurements provides the best match. 
Figure~\ref{fig:simspec} shows an example simulated quasar spectrum
from our models, and compares the Ly$\alpha$ forest observables
(mfp and $\tau_{\rm eff}$) obtained from our simulated forest spectra to 
observations of high redshift quasars.

\subsection{Color Selection}\label{sec:color_cuts}

Based on the results from the simulated spectra, 
we concluded that the most efficient use of telescope time
was to focus on a fairly narrow redshift range, $4.7 < z < 5.1$,
where the colors of stars and quasars are best separated and
the selection efficiency is high. 
Figure~\ref{fig:color_color} shows the $gri$, $riz$, and $riJ$ 
colors of red stars and high redshift quasars in our sample, as well as
the selection criteria we adopted:
\begin{itemize}
\item $i < 22.0$
\item $S/N(u) < 2.5$
\item $g-r > 1.8$ OR $S/N(g)<3.0$
\item $r-i > 1.2$
\item $i-z < 0.625((r-i)-1.0)$
\item $i-z < 0.55$
\item $i-J < ((r-i)-1.0) + 0.56$ 
\end{itemize}
where all magnitudes are in the AB system, measured within fixed apertures
(as described in \S\ref{sec:imaging_data}), and have been corrected for
Galactic extinction using the maps of \citet{SFD98}. $S/N$ measurements
are obtained from the aperture fluxes and uncertainties given by
Sextractor.

We apply no morphological criteria in our candidate selection
after finding that the number of potentially resolved objects
(Sextractor CLASS\_STAR $< 0.8$) is small ($<10\%$ of the total selected),
indicating that contamination from compact galaxies is negligible.

These criteria are similar but not identical to the cuts that 
define the $z>4.5$ inclusion region in the SDSS quasar selection 
algorithm, as given in \citet{Richards+02sel}. 
We extend the selection
nearly 2 mag fainter than the SDSS main survey by utilizing
the coadded imaging on Stripe 82. We define dropout criteria
in the $u$ and $g$ bands through $S/N$ criteria within fixed apertures
rather than a magnitude cut. Although the Stripe 82 coadded imaging
is highly uniform (e.g., Fig.~\ref{fig:stripe82map}), using a $S/N$
threshold more fully utilizes the $u$-band depth at a given location,
as $z\sim5$ quasars are not expected to have any detectable flux in
this band. We bracket
the redshift range of the search to $z\ga4.7$ by
requiring a red $r-i$ color and to $z\la5.1$ by requiring
a blue $i-z$ color. Finally, we take advantage of the available
infrared data from UKIDSS, effectively imposing a veto on
objects that are too red in $i-J$. This cut is made using the
aperture flux ratios in the $i$ and $J$ bands, so that objects undetected 
in the UKIDSS $J$ band --- as expected for $z\sim5$ quasars --- pass
the final cut without imposing a $S/N$ threshold.

The $i$-band detection catalog contains 1.5M objects. We apply
loose pre-selection cuts in the $riz$ colors
(see middle panel of Fig.~\ref{fig:color_color} and \S\ref{sec:UKIDSS_imaging})
to reduce this list to $\sim10,000$ 
potential candidates for which UKIDSS images are downloaded
and analyzed. 
The near-IR photometry is highly useful as stellar veto:
of the $\sim10,000$ pre-selected candidates from the
optical imaging, only $\sim650$ are rejected by the $i-J$ color
cut we adopt (\S \ref{sec:color_cuts}) using the UKIDSS catalog photometry, while $\sim4500$ are rejected based on the aperture fluxes.
After applying the final color criteria listed above, we have 92 candidates 
to a limit of $i_{\rm AB}=22$.

\section{Spectroscopic Observations and Data Reduction}\label{sec:specobs}

We obtain spectroscopic identifications of our candidates from
multiple sources by utilizing the dense SDSS/BOSS spectroscopic 
coverage of Stripe 82 combined with our own observations. First, we 
identify a few of the brightest candidates on Stripe 82 using 
spectroscopic data from the SDSS I/II \citep{dr7}. More recently, 
the BOSS collected a large number of spectra on Stripe 82, including 
several ancillary science programs \citep{Dawson+12} targeting high 
redshift quasars. Finally, we have obtained spectra for additional
candidates (extending much fainter than the SDSS and BOSS spectroscopic 
observations) using the MMT and Magellan telescopes.
In total, we have 73 spectroscopic identifications for our 92 
candidates (79\%). The candidates with spectroscopic observations can be
considered an unbiased subset of the full candidate sample, as will be
discussed further in \S~\ref{sec:completeness}.
Table~\ref{tab:summary} summarizes the status 
of spectroscopic observations of $z\sim5$ quasar candidates on Stripe 
82, including a number of confirmed quasars not within the uniform sample.

\subsection{SDSS DR7}\label{sec:dr7spec}

The SDSS I/II spectroscopic survey concluded with the DR7 release 
\citep{dr7}. Spectra were reduced with the standard SDSS pipeline 
\citep{Stoughton+02}. Quasar target selection is described in 
\citet{Richards+02sel}; high-redshift quasars are targeted through 
two methods, both to a limit of $i=20.2$. First, outliers from the 
stellar locus in $griz$ space are identified as high-redshift quasar 
candidates, and second, various color cuts aim to extend the 
quasar yield within specific redshift ranges beyond the locus
outlier selection. \citet{dr7qso} present a catalog of over 
100,000 spectroscopically confirmed quasars drawn from 
$\approx9380~\rm{deg}^2$ of the DR7 spectroscopic footprint (hereafter 
DR7QSO). This catalog includes 191 quasars at $4.7 < z < 5.1$ to a limit 
of $i \la 20.2$, of which 184 are selected by our color criteria based
on their SDSS flux measurements.

We utilize the DR7 data in several ways. First, we matched our 
candidate list to the DR7QSO catalog, as well as to the full DR7 
spectroscopic database. We identified eight of our candidates as 
confirmed $4.7 < z < 5.1$ quasars in DR7QSO. There were three additional
quasars in this redshift range that did not meet our color criteria.
None of our candidates matched to non-quasars among the 1.6 million
objects with spectra in DR7.

We further employ the DR7 quasars to determine the bright end of the 
luminosity function, by constructing a uniform sample of DR7QSO quasars in 
our redshift range drawn from the SDSS DR7 imaging footprint. We will 
describe this sample in \S\ref{sec:dr7complete}.

\subsection{BOSS}\label{sec:boss_spec}

The BOSS quasar survey is primarily designed to find quasars at 
$2.2 < z < 3.5$ for the purpose of Ly$\alpha$ forest studies 
\citep{ME07,Ross+12}. The Extreme Deconvolution algorithm (XDQSO) used 
as the primary quasar targeting method in BOSS is tuned to
this redshift range and is highly efficient \citep{Bovy+11}.
However, a large section of Stripe 82 was observed by BOSS in Fall 2010 as
BOSS Chunk 11 \citep{Ross+12}; during these observations quasars
were targeted through a combination of color selection (using XDQSO)
and variability selection \citep{PD11}. The highest
redshift quasar identified by the main BOSS targeting in Stripe 82,
including variability, is at $z=4.46$. 

In addition, we are conducting ancillary science programs in BOSS that 
take advantage of unused fibers once the primary survey targets (galaxies
and mid-$z$ quasars) have been allocated fibers \citep{Dawson+12}.
These programs extend the $gri$ and $riz$ color cuts described
in \citet{Richards+02sel} to fainter quasars by utilizing areas with 
multiple epochs of imaging in SDSS I/II due to overlapping or repeated 
scans. On Stripe 82 these ancillary programs utilized photometry from 
the $\sim20$-epoch coadded images available in the DR7 CAS and 
described in \citet{Annis+11};
these observations are included in BOSS DR9 \citep{dr9}. We will
analyze the quasars identified in overlap regions outside of Stripe 82
in future work.

The selection criteria for $z>4.6$ quasars are similar to the $riz$ 
inclusion region in \citet{Richards+02sel}, but extended to a faint 
limit of $i=21.5$ and using the catalogs from the \citet{Annis+11}
coadded imaging for target selection.
On Stripe 82, this program was allocated 283
targets, of which 221 received spectra, but only 22 were high-$z$
quasars. Of those, 10 were previously known from SDSS DR7; thus
the program yielded only 12 new quasars. All 12 meet our selection
criteria and are included in this study. The ancillary program targets 
that are not quasars fall mainly into two categories:
1) objects with fluxes contaminated by nearby bright stars, and
2) spurious or moving objects (e.g., asteroids). 
We matched the failed targets to the coadded image catalogs and found that
either they did not have matches in our coadded imaging, or the
matches did not meet our selection criteria.

BOSS spectra are collected with a fiber-fed, multi-object spectrograph
\citep{Smee+12} and reduced with a pipeline described in 
\citet{Bolton+12}. We visually  examined all of the spectra for 
ancillary program targets on Stripe 82,  as well as any classified by 
the pipeline as having $z>3.6$. We also cross-checked our examinations 
against the DR9 Quasar Catalog as described in \citet{dr9qso}. As with 
SDSS, we cross-checked our candidate list against all spectra in BOSS, 
not just confirmed quasars, again finding no matches to non-quasars.
Figure~\ref{fig:bossspec} displays the BOSS spectra of $z>4.7$
quasars on Stripe 82.

\begin{deluxetable}{lrr}
 \centering
 \tablecaption{Status of high redshift quasar candidates on Stripe 82}
 \tablewidth{0pt}
 \tablehead{
  \colhead{Sample} &
  \colhead{Uniform} &
  \colhead{All} 
 }
 \startdata
    Candidates & 92 & - \\
    ... spec. ids & 73 & 106 \\
    ... quasars & 71 & 84 \\
    ... $4.7<z<5.1$ & 52 & 64 
 \enddata\label{tab:summary}
\end{deluxetable}

\subsection{Magellan observations}\label{sec:magellan}

The remaining candidates on Stripe 82 ($\sim70$) were mostly fainter 
than $i=21$ and required spectroscopic observations on larger 
telescopes. We first observed four candidates at the Magellan Clay
6.5m on 2011 Jun 11-13 using the Magellan Echellette spectrograph 
\citep[MAGE;][]{mage}. MAGE provides full coverage from 3100\AA\ to 
1$\mu$m at a resolution of $\sim5800$ using the 0.7\arcsec\ slit. The 
typical seeing during this run was 0.6\arcsec.

Magellan spectra were reduced using MASE \citep{mase}, an IDL-based
pipeline designed for MAGE data. Wavelength calibration was provided 
by ThAr lamps observed shortly after the targets and at a similar 
airmass, and the standard star Feige 110 was used for flux calibration.
Of the four targets observed at Magellan, three are $z\sim5$
quasars (Figure~\ref{fig:magespec}; one object had been previously 
observed in BOSS), and one is a star. 

\subsection{MMT observations}\label{sec:MMT}

The bulk of our spectroscopic observations occurred at the MMT 6.5m 
telescope using the Red Channel spectrograph. We used the 270~mm$^{-1}$ 
grating centered at 7500~\AA\footnote{Some objects were observed at
a redder setting during the course of a program targeting higher redshift 
objects.}, providing coverage from 5500\AA\ to 9700\AA. We used either 
the 1\arcsec\ or 1.5\arcsec\ slit based on the seeing, providing 
resolutions of $R\sim640$ and $R\sim430$, respectively.

Observations were conducted on 2011 Jun 23-24, 2011 Oct 1-4, and 
2012 May 27-28. During 2011 June nine objects were observed in poor 
seeing (1.5-3\arcsec); five were confirmed as quasars. Conditions during 
the 2011 October run were fair with $\la1$\arcsec\ seeing; however, most 
of the run was lost to thick clouds and high humidity, particularly 
during the early part of the night. As a result, we obtained few spectra 
for candidates at $20^{\rm h} < {\rm R.A.} < 22^{\rm h}$. In total, 52 
targets were observed in October, of which 41 were confirmed quasars. 
In the 2012 May run, 17 additional candidates with 
$20^{\rm h} < {\rm R.A.} < 24^{\rm h}$ were observed, resulting in 
11 new high-redshift quasars. Conditions during this run were excellent, 
with $0.7\mbox{--}1.2$\arcsec\ seeing throughout. In general we valued 
efficiency over quality. Therefore the exposure times were short, between 
5 and 15 minutes with typically a single exposure per target, and many of the 
spectra have a low signal-to-noise ratio. 
However, the quasars among the observed targets are easily identified 
by their prominent emission lines, while the few contaminants
can be ruled out as high-redshift quasars by the lack of either emission
lines or a strong spectral break towards blue wavelengths.

Data were processed using standard longslit reduction techniques
through a combination of Pyraf\footnote{Pyraf is a product of the 
Space Telescope Science Institute, which is operated by AURA for NASA.}
and python routines, including bias subtraction, pixel level corrections 
from flat fields generated from internal lamps, and sky subtraction using 
a polynomial background fit along the slit direction. Cosmic rays were
idntified and masked using the LACOS routines \citep{lacos}.  Wavelength 
calibration was provided from an internal HeNeAr lamp, and then corrected 
on a per-image basis using night sky lines (primarily the OH line list 
given by \citealt{Rousselot}). Standard stars were observed between one 
and three times per night and used for flux calibration. However, the
conditions were highly variable and the absolute flux calibration
of the spectra is not reliable. 
Figure ~\ref{fig:mmtspec_1} shows the MMT spectra obtained for 
high redshift quasars on Stripe 82.

\section{Quasar catalog}\label{sec:catalog}

Table~\ref{tab:quasars} consists of our full Stripe 82 $z\sim5$ quasar 
catalog, consisting of:
\begin{itemize}
\item 11 SDSS DR7 quasars with $z \ge 4.7$,
\item 14 BOSS DR9 quasars with $z \ge 4.7$,
\item 59 quasars with spectra obtained at MMT and Magellan.
\end{itemize}

For all objects, we provide photometry in the $griz$ bands derived
from the coadded imaging that formed the basis for our target
selection. We also provide $J$-band photometry obtained from 
our aperture photometry of the UKIDSS DR8plus images, or,
when available, imaging from MMT SWIRC.
The table includes all $z>4.7$ quasars on Stripe 82 from 
the three data sources, but not all of these quasars are included
in the uniform sample used to calculate the QLF (\S\ref{sec:results}). 
In \S\ref{sec:s82complete} we will derive the selection probability
for each quasar, this value is included in the catalog and we flag
quasars that are not part of the uniform sample by assigning them
a value of $-1$.

The only line widely available at a reasonable $S/N$ in our spectra 
is the Ly$\alpha$ line. Our spectra do generally cover the \ion{C}{4}
emission region; however, this line shows offsets from the systemic 
redshift that are correlated with properties such as luminosity and 
radio loudness \cite[e.g.,][]{Richards+11}. Therefore, we assign 
redshifts based on a combination of fitting the Ly$\alpha$ 
line\footnote{The Ly$\alpha$ line can also have systematic offsets, 
in particular a $\sim500~{\rm km}~{\rm s}^{-1}$ redshift due to 
absorption of the blue wing by the Ly$\alpha$ forest 
\citep[e.g.,][]{Shen+07}. We do not account for this offset as it
is much smaller than the uncertainties we assume for our redshift
designations.}, measuring the onset of Ly$\alpha$ forest absorption, 
and visually matching a quasar template spectrum to the observed spectra. 
In general, the redshifts have an uncertainty of $\Delta z \sim 0.02$, 
which is sufficiently accurate for calculation of a luminosity function.
All of our quasars are consistent with a Type I classification based on their
broad line widths ($FWHM>1000~{\rm km}~{\rm s}^{-1}$).

\begin{figure}[!t]
 \epsscale{1.1}
 \plotone{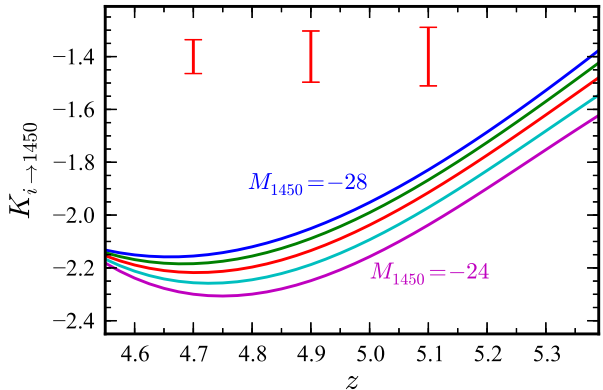}
 \caption{Luminosity-dependent $k$-correction as determined from
 the quasar spectral simulations.
 The lines show the $k$-correction from observed $i$-band magnitude
 to $M_{1450}$ for $M_{1450}=-28$ (top) to $M_{1450}=-24$ (bottom),
 in steps of $\Delta M=1$.
 The trend with redshift is fairly smooth; although Ly$\alpha$ enters
 the $i$ band at $z\approx4.6$, Ly$\alpha$ forest absorption tends
 to offset its effect on the $k$-correction.
 The vertical error bars at the top of the plot denote the representative
 scatter in the $k$-correction at the median luminosity ($M_{1450}=-26$)
 at three redshifts.
 This scatter is also determined from the simulations and accounts for the
 variations in emission line strengths, continuum shapes, and Ly$\alpha$
 forest absorption included in the quasar spectral model.
 These corrections are from the Stripe 82 simulations; the DR7 simulations
 produce similar results but are adjusted for the asinh magnitude scale.
 \label{fig:kcorr}
 }
\end{figure}

We use the simulated spectra derived from our quasar model 
(\S\ref{sec:simulations}) to derive $k$-corrections as a function
of both redshift and luminosity. 
We derive the average correction for the observed $i$-band flux
to the monochromatic luminosity at rest-frame 1450~\AA~($M_{1450}$)
for a large number of simulated
quasars in narrow bins of $(M,z)$ (see \S\ref{sec:completeness}), then interpolate
this grid to derive an individual quasar correction.
Figure~\ref{fig:kcorr} shows the luminosity-dependent $k$-corrections
derived from our quasar model and used in this work.
Accounting for the average emission line contribution to the $k$-correction
as a function of luminosity alleviates some of the
issues arising from the fact that our best photometry is in
the $i$-band, which contains the Ly$\alpha$ line.\footnote{Note
that the $i$-band measurement is from the coadded imaging, and is
thus an average of a decade of individual measurements, smoothing
over the variable lightcurve of each quasar.}
This approach also corrects for some of the bias introduced by 
luminosity-dependent line emission and its effect on
broadband photometric data.
However, the distribution of intrinsic quasar SEDs at a given
luminosity and redshift is quite broad, introducing scatter
into our absolute magnitude calculations (also shown in
Figure~\ref{fig:kcorr}).

Alternatively, the $k$-correction can be derived directly from the
spectral data \citep[see, e.g.,][for a discussion of spectral
$k$-corrections for high$-z$ quasars]{Glikman+11}. However, we are
again limited by the low $S/N$ in the continuum of our spectra.
In addition, we do not attempt an accurate flux calibration of our
spectra (indeed, the BOSS quasar spectra are known to have flux
calibration errors in some instances, see \citealt{dr9qso}).
Thus we would need to calibrate the observed (noisy) spectra with
the broadband fluxes from the imaging. We chose to use a template
$k$-correction based on photometry as it can be more consistently applied.

\begin{figure}[!t]
 \epsscale{1.1}
 \plotone{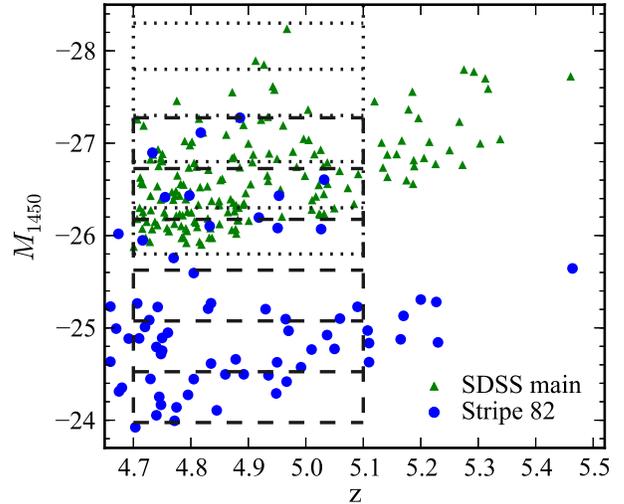}
 \caption{Distribution of high-$z$ quasars in our sample in
 luminosity and redshift. Green triangles represent DR7 quasars, 
 blue circles denote Stripe 82 quasars. The grid lines show the
 bin edges used for the calculation of the binned QLF, with
 dashed lines for Stripe 82 and dotted lines for DR7.
 \label{fig:Mz}
 }
\end{figure}

Figure~\ref{fig:Mz} shows the distribution in redshift and luminosity for
both the Stripe 82 and SDSS main samples, after applying our $k$-corrections.

\subsection{Notes on individual objects}\label{sec:objnotes}

In this section, we note objects that have radio detections,
uncertain identifications from the spectroscopy, and unusual
spectral features.

{\bf J221941.90+001256.2 ($z=4.30$), 
     J224524.27+002414.2 ($z=5.16$)}:
 These two objects have radio counterparts at 1.4~GHz from VLA imaging of
 Stripe 82 \citep{Hodge+11}. They are the only sources with counterparts
 in that catalog, which is derived from imaging over 92~deg$^2$ to a depth
 of 52~$\mu$Jy~beam$^{-1}$. 
 J221941.90+001256.2 has a peak flux density of 
 $F_{\rm 1.4,pk}=0.92\pm0.07$~mJy~beam$^{-1}$ and 
 J224524.27+002414.2 has $F_{\rm 1.4,pk}=1.09\pm0.06$~mJy~beam$^{-1}$.
 Both sources also have counterparts in the Faint Images of the Radio
 Sky at Twenty-Centimeters (FIRST) catalog \citep{BWH95}, with peak
 flux densities of $F_{\rm 1.4,pk}=0.87\pm0.10$~mJy~beam$^{-1}$
 and $F_{\rm 1.4,pk}=0.92\pm0.10$~mJy~beam$^{-1}$, respectively.
 Neither is included in our uniform sample as they lie outside the 
 defined redshift range. J221941.90+001256.2 also shows strong
 broad absorption line (BAL) features.

{\bf J223907.56+003022.6 ($z=5.09$), 
     J232741.35-002803.9 ($z=4.75$),
     J021043.16-001818.4 ($z=5.05$)}:
 These three objects also have FIRST counterparts, with peak
 flux densities of $F_{\rm 1.4,pk}=1.35\pm0.10$~mJy~beam$^{-1}$,
 $F_{\rm 1.4,pk}=1.24\pm0.12$~mJy~beam$^{-1}$, and
 $F_{\rm 1.4,pk}=6.82\pm0.10$~mJy~beam$^{-1}$, respectively.
 All three are included in the uniform sample.

{\bf J211158.01+005302.6 ($z=4.98$), 
     J234730.56+002306.3 ($z=4.71$)}: 
 We identified these objects as quasars based on their discovery 
 spectra; however, the spectra are noisy and the identifications 
 were uncertain. We later confirmed them as quasars with MMT 
 observations on 2012 Aug 25.

{\bf J211225.39-000141.3 ($z=4.67$), 
     J030315.05-000347.6 ($z=4.72$)}:
 These spectra also have low $S/N$. Both objects appear to have a 
 Lyman break feature (more evident in the 2D spectra) and Ly$\alpha$ 
 and \ion{N}{5} emission features, with apparent \ion{N}{5} 
 absorption troughs. We include these objects as confirmed quasars.
 J030315.05-000347.6 may have BAL features.

{\bf J222018.48-010146.8 ($z=5.62$), 
     J000552.33-000655.6 ($z=5.86$)}:
 These two quasars were targeted as part of an SDSS+UKIDSS high-$z$
 BOSS quasar ancillary program. J000552.33-000655.6 was first
 reported in \citet{Fan+04PIII}.

\section{Results}\label{sec:results}

\begin{figure*}[!t]
 \epsscale{1.1}
 \plotone{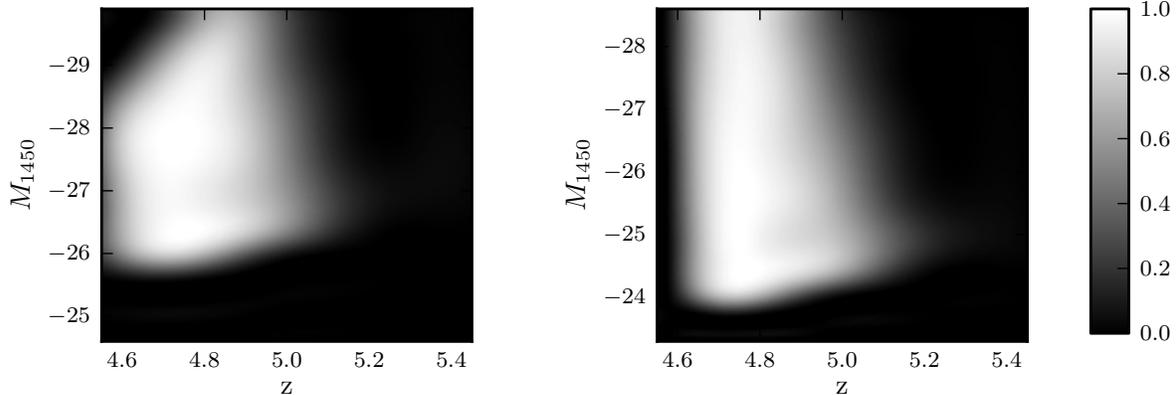}
 \caption{
 Selection functions for the SDSS main sample (left, $i_{\rm asinh}<20.2$) 
 and Stripe 82 (right, $i_{\rm AB}<22$). Note the difference in scale on the 
 $y$-axis, as the Stripe 82 survey reaches nearly two magnitudes fainter than
 the SDSS main.
 The selection probability is calculated in each $(M,z)$ bin by determining
 the fraction of the 200 simulated quasars in each bin that pass the selection
 criteria, averaging over the SED distribution.
 \label{fig:selection_fun}
 }
\end{figure*}
\begin{figure}[!t]
 \epsscale{1.1}
 \plotone{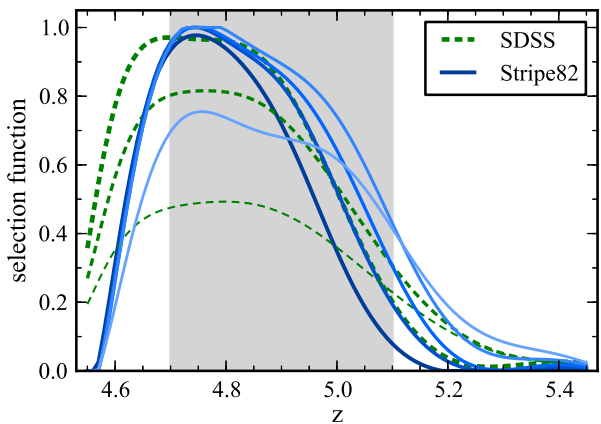}
 \caption{
 Comparison of the selection functions for the SDSS main sample 
 (green) and Stripe 82 (blue). The lines represent slices at fixed observed
 magnitude through the functions shown in Figure~\ref{fig:selection_fun}.
 For the SDSS main, the slices are at $i_{\rm asinh}=(19,20,20.2)$, and for 
 Stripe 82 they are at $i_{\rm AB}=(19.1,20.2,21,21.5,21.9)$. The lines
 become thinner and lighter with increasing magnitude.
 We restrict our uniform sample to $4.7<z<5.1$
 where the completeness is highest (gray shaded region). Although 
 the completeness is relatively high at $z\sim4.65$, the selection
 function is steep here, which would make incompleteness corrections
 very sensitive to systematics in the selection function model.
 \label{fig:selection_compare}
 }
\end{figure}
\begin{figure}[!ht]
 \epsscale{1.1}
 \plotone{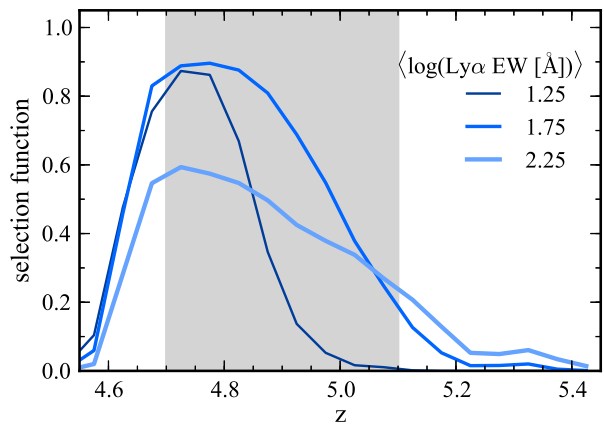}
 \caption{
 Dependence of the Stripe 82 selection function value derived 
 from the simulations on the Ly$\alpha$ equivalent width. 
 The simulated quasars are divided into three bins of Ly$\alpha$ EW, with
 median values shown on the plot.
 The anticorrelation between EW and 
 luminosity means that objects with greater Ly$\alpha$ EW are fainter
 on average, and thus have a lower selection probability. On the other hand,
 the Ly$\alpha$ line itself introduces redshift-dependent trends. For example,
 objects at $z\ga5$ with strong Ly$\alpha$ emission will have redder 
 $r-i$ and bluer $i-z$ 
 colors, and thus be more likely to pass our selection criteria.
 This effect is apparent in Fig.~\ref{fig:color_color}, which shows that
 lower luminosity sources fall within the color selection criteria over a 
 wider redshift range than higher luminosity sources.
 \label{fig:selection_lya}
 }
\end{figure}

\subsection{Survey Completeness}\label{sec:completeness}

We use the simulations described in section~\ref{sec:simulations} to 
estimate the completeness of our selection criteria. To derive a selection 
function, we construct a grid of simulated quasars distributed evenly in 
($M_{1450},~z$) space with 200 quasars per bin of 
$\Delta M = 0.1$, $\Delta z = 0.05$. Each object is sampled from
models for quasar emission properties as outlined in
\S~\ref{sec:simulations}. 
The simulated quasars are then passed through the color cuts after
adding photometric errors,
and the fraction of objects within each bin that pass the cuts provides
an estimate of the completeness for a given luminosity and redshift,
under the assumption that our quasar model accurately represents
the intrinsic distributions of quasar properties.

We generate two simulation grids. The first adopts photometric uncertainties 
typical of the SDSS main survey and uses asinh magnitudes; we use this 
simulation to derive the completeness of the SDSS quasar survey at $z\sim5$. 
The second uses photometric uncertainties from the coadded imaging and is 
used for the Stripe 82 sample. The uncertainties are determined by fitting
a relation to the uncertainties of stellar objects in the Stripe 82 catalog
as a function of band flux.

\subsubsection{DR7 Completeness}\label{sec:dr7complete}

The SDSS quasar selection algorithm \citep{Richards+02sel} was not 
finalized until the survey was already in progress, thus to construct 
a statistical sample from DR7 we follow the method given in 
\citet{Richards+06LF} to identify DR7 quasars from regions with
uniform target selection. This limits the final area to 
$6222~{\rm deg}^2$. We further cut the DR7 sample to only objects 
selected by the $riz$ color inclusion regions (excluding objects selected 
only as  $griz$ stellar locus outliers); this results in the loss of only 
a few objects, but greatly simplifies the calculation of the selection 
function. After restricting to the uniform targeting area and applying 
the color cuts, DR7QSO provides 146 quasars with 
$4.7<z<5.1$\footnote{We do not double-count DR7 quasars that lie on 
Stripe 82 when calculating the LF, as the uniform targeting area does 
not include Stripe 82.}.

We use the simulated quasar photometry combined with the $riz$ color 
cuts of \citet{Richards+02sel} to derive the selection function for 
$z\sim5$ quasars in the main SDSS sample (Figure~\ref{fig:selection_fun}). 
We add a 5\% correction for photometric incompleteness (i.e., objects 
lost due to crowding or proximity to bright stars or galaxies, see 
\citealt{Richards+06LF}) and spectroscopic incompleteness of 5\% 
(objects selected by the targeting algorithm but without a spectrum in 
DR7). We determined the latter incompleteness by querying the DR7 
Target database for objects with the QSO\_HIZ flag, finding that 5\% 
lack spectra. The missing spectra are mainly due to fiber collisions 
that result when tiling spectroscopic targets, due to the restriction 
that fibers on a single plate must be separated by $>55$\arcsec.

\begin{figure}[!t]
 \epsscale{1.15}
 \plotone{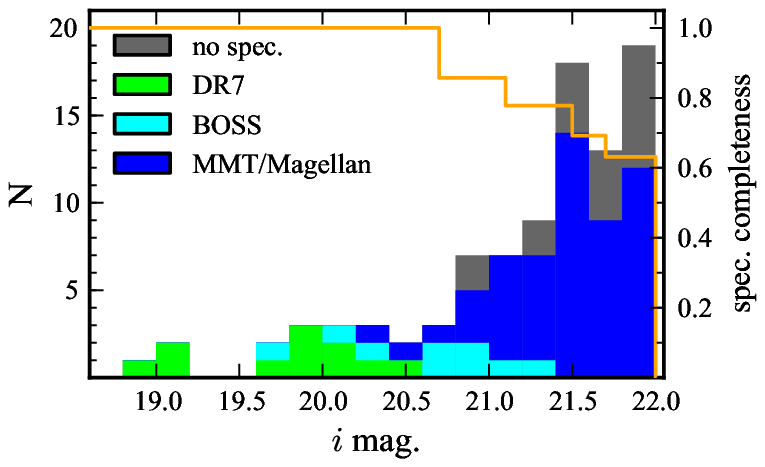}
 \caption{Spectroscopic completeness of $z\sim5$ quasar candidates from
  the uniform sample on Stripe 82 ($i_{\rm AB}<22$). The histogram is 
  divided into objects without spectroscopic observations (gray), and 
  those with spectra obtained through DR7 (green), BOSS (cyan), and 
  MMT/Magellan observations (blue). The orange line represents the
  spectroscopic incompleteness correction applied when calculating the
  QLF in order to account for candidates without spectra.
 \label{fig:spec_complete}
 }
\end{figure}

\subsubsection{Stripe 82 Completeness}\label{sec:s82complete}

Although the Stripe 82 selection function is obtained from the same quasar
model as for DR7, it is evident from Figure~\ref{fig:selection_fun} that the
Stripe 82 selection probes a different population.
First, the Stripe 82 data are much deeper and thus the completeness
remains high to lower luminosities. Also, the color selection criteria
are somewhat different.
Figure~\ref{fig:selection_compare}
compares the selection efficiencies for DR7 and Stripe 82 as a function
of redshift.
It is noteworthy that the selection function value is higher for fainter
objects at some redshifts. This is because the selection function
calculation averages over a range of colors in each ($M,z$) bin, with
the colors depending on the continuum slopes, line strengths, and other
features sampled from the model. At $z\sim5$, our color selection
is quite sensitive to the Ly$\alpha$ equivalent width, as shown in
Figure~\ref{fig:selection_lya}. Because the EW increases at lower
luminosities and Ly$\alpha$ is fully within the $i$-band in our redshift
range, the colors of fainter objects are redder in $r-i$ and bluer in
$i-z$ than brighter objects, increasing their likelihood of meeting our
selection criteria (see also Figure~\ref{fig:color_color}). This
effect is not captured by models that do not account for the
Baldwin Effect.

Our selection function models generally predict near-zero completeness for
$z\ga5.2$ quasars; however, it is clear from Figure~\ref{fig:Mz} that our
color criteria do select a non-negligible number of quasars at these 
redshifts (particularly for the DR7 data). We found that models with
increases in both the mean and scatter of the Ly$\alpha$ EW distribution
provide a better fit to the observed colors (and the $z>5.2$ redshift
distribution) than our fiducial model. However, these models had little 
effect on our primary analysis, as we restrict to the range $4.7 < z < 5.1$,
where our fiducial model already predicts high completeness. We thus chose
to maintain consistency with the BOSS analysis presented in 
\citet{Ross+12qlf} and adopt the same quasar spectral model.

We estimate the photometric completeness to be 95\% 
\citep[see, e.g, Fig.~7~of ][]{Annis+11}.
The spectroscopic coverage of the Stripe 82 candidates is high: 
$\sim90\%$ of candidates with $i_{\rm AB}<21.5$ have spectra. 
Near the faint limit, the completeness is a bit lower: 69\% of 
candidates with $21.5<i_{\rm AB}<21.7$ and 63\% of candidates with
$21.7<i_{\rm AB}<22$ have spectra. Figure~\ref{fig:spec_complete} 
shows our spectroscopic completeness as a function of observed magnitude,
as well as the correction we use to account for the missing spectra in our
luminosity function calculation. In total, 19 candidates do not
have spectroscopic identifications; by applying this correction we assume
that they are a random subsample of the full candidate set. This
assumption is fair given that the choice of which targets to observe
was largely constrained by sky location and weather conditions, not
properties of the candidates themselves (such as color).

The Stripe 82 survey includes
52 quasars at $4.7 < z < 5.1$. The Stripe 82 sample is not only
highly complete but also highly pure: out of 73 candidates with
spectroscopic identifications, 71 are high-redshift quasars ($z>4$), and
52 (71\%) are quasars in the targeted redshift range. For completeness,
we provide in Table~\ref{tab:targets_noid} a list of the candidates
that were either not observed spectroscopically (19 objects), or were 
found not to be high-redshift quasars (2 objects, both have no emission
features and appear to be stellar continua).

\subsection{Binned Luminosity Function}\label{sec:binnedqlf}

We first calculate the luminosity function from the combined SDSS 
main and Stripe 82 $z\sim5$ quasar samples by dividing the sample
into discrete bins of luminosity and redshift. Guided by our 
completeness calculations, we restrict the sample to the interval
$4.7 < z < 5.1$, where the selection efficiency is relatively high. 
We refer to this as the uniform sample. We  use a single redshift bin, 
ignoring any evolution of the QLF parameters over the width of the bin. 
In particular, the redshift evolution derived by \citet{Fan+01LF} predicts 
a decline by a factor of $\sim1.5$ in space density from $z=4.7$ to $z=5.1$.
We do not account for this evolution in the binned QLF, but it will
be incorporated below in a maximum likelihood fit to each quasar.
We calculate the binned luminosity function using the $1/V_a$ method
\citep{Schmidt68,AB80}, including the correction of \citet{PC00}. The calculation
is performed separately on the DR7 and Stripe 82 data, accounting for
the differences in sky area, depth, and selection criteria between the
two samples.

\begin{figure}[!t]
 \epsscale{1.1}
 \plotone{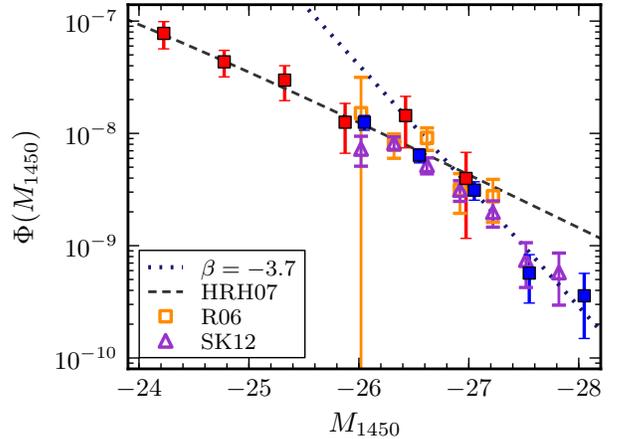}
 \caption{
 Binned QLF at $\langle{z}\rangle=4.9$ from SDSS main (blue) and 
 Stripe 82 (red).
 The dotted line shows a single power law fit to points with 
 $M_{1450}<-27.0$ ($\beta=-3.7$). The departure from a single power
 law is evident when the faint Stripe 82 data are included.
 For comparison, we show previous calculations of the binned QLF
 at $z=4.75$, corrected for our cosmology and shifted to $z=4.9$
 using the redshift evolution from \citet{Fan+01LF}.
 The orange points represent the SDSS DR3 calculation from an
 area of 1622~deg$^2$ \citep{Richards+06LF}, while the purple points
 show the SDSS DR7 calculation from \citet{SK12} using a sample nearly
 identical to our SDSS main sample covering $\sim6200~{\rm deg}^2$.
 The dashed lines correspond to the \citet{HRH07} QLF model discussed
 in \S\ref{sec:qlfevol}; most of the constraint on this model at $z=5$ comes
 from the \citet{Richards+06LF} points.
 \label{fig:QLF_comparedata}
 }
\end{figure}

Table~\ref{tab:binnedqlf} provides the binned QLF for both the SDSS main
(DR7) and Stripe 82 samples. We include the number counts in each 
magnitude bin, as well as the corrected number counts after accounting
for all sources of incompleteness. The binned QLF data is also displayed in
Figure~\ref{fig:QLF_comparedata}. The SDSS data show a steep
drop in the number density at the bright end; from the combined
data it is evident that the QLF becomes shallower towards lower 
luminosities. Fitting a single power law to the binned data from both surveys 
at the bright end ($M_{1450}<-27.0$) results in a steep slope of $\beta=-3.7$.

\citet{SK12} have also calculated the binned QLF of SDSS
quasars at $z=4.75$, repeating the DR3 analysis of 
\citet{Richards+06LF} for the larger DR7 sample. Our
methodology differs from that of \citet{SK12}
in several respects. 
Most notably, we have recalculated the selection function
and $k$-corrections using our new quasar model. In addition,
we restrict the quasar sample  to color-selected
objects (excluding those identified only as stellar locus outliers).
The \citet{Richards+06LF} selection function has a value
of $\sim1$ at $z\ga5$, much higher than the values we obtain 
(see Fig.~\ref{fig:selection_compare}), and inconsistent with our
previous finding (\S\ref{sec:dr7complete}) that few quasars were
selected by locus outlier criteria alone. However, although our selection
function disagrees with that of \citet{Richards+06LF} by as much as
a factor of $\sim2$ at $z\sim5$, the highest redshift bin ($4.5<z<5.0$) 
in both \citet{Richards+06LF} and \citet{SK12} is dominated by objects
near the low redshift edge of the bin.
Finally, we use a slightly different method for calculating the
spatial area of the survey; however, we obtain a similar
result ($6222~{\rm deg}^2$ vs. $6248~{\rm deg}^2$). 
Figure~\ref{fig:QLF_comparedata} compares our binned QLF to 
\citet{Richards+06LF} and \citet{SK12}; the agreement is generally
good, though differences of $\sim20\mbox{--}30$\% may still be
attributed to the different approaches used.

\begin{deluxetable}{rrrrr}
 \centering
 \tablecaption{Binned QLF}
 \tablewidth{2.25in}
 \tablehead{
  \colhead{$M_{1450}$} &
  \colhead{$N$} &
  \colhead{$N_{\rm cor}$} &
  \colhead{$\log\Phi$\tablenotemark{a}} &
  \colhead{$\sigma_\Phi$\tablenotemark{b}} 
 }
 \startdata
\multicolumn{5}{c}{DR7} \\
-28.05 &  3 &   4.8 & -9.45 & 0.21 \\
-27.55 &  5 &   7.7 & -9.24 & 0.26 \\
-27.05 & 30 &  42.0 & -8.51 & 0.58 \\
-26.55 & 57 &  85.1 & -8.20 & 0.92 \\
-26.05 & 51 &  81.5 & -7.90 & 1.89 \\[6pt]
\multicolumn{5}{c}{Stripe 82} \\
-27.00 &  2 &   2.2 & -8.40 & 2.81 \\
-26.45 &  5 &   8.1 & -7.84 & 6.97 \\
-25.90 &  5 &   7.0 & -7.90 & 5.92 \\
-25.35 & 10 &  16.7 & -7.53 & 10.23 \\
-24.80 & 15 &  24.3 & -7.36 & 11.51 \\
-24.25 & 14 &  26.8 & -7.14 & 19.90 
 \enddata
\label{tab:binnedqlf}
 \tablenotetext{a}{$\Phi$ is in units of Mpc$^{-3}$~mag$^{-1}$.}
 \tablenotetext{b}{$\sigma_\Phi$ is in units of $10^{-9}$~Mpc$^{-3}$~mag$^{-1}$.}
\end{deluxetable}

\subsection{Parameter Estimation From Maximum Likelihood}\label{sec:mlefit}

\begin{figure}[!t]
 \epsscale{1.1}
\ \plotone{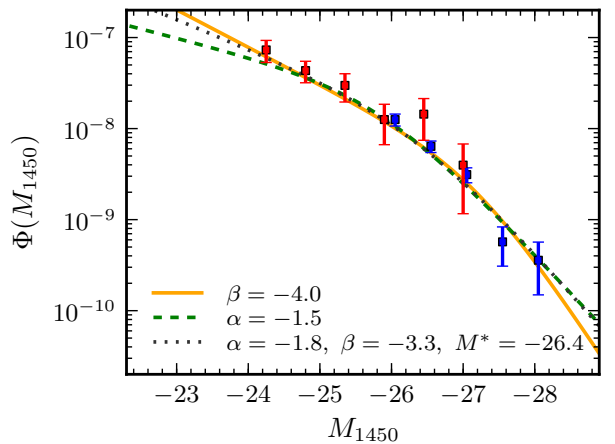}
 \caption{
 The double power law fits from maximum likelihood analysis 
 (Table~\ref{tab:mlefit}) plotted with the binned QLF data. Each model fit 
 is labeled by the choice of fixed parameters for that fit.
 \label{fig:QLF_fits}
 }
\end{figure}
\begin{deluxetable}{cccc}
 \centering
 \tablecaption{MLE fit parameters}
 \tablewidth{0pt}
 \tablehead{
  \colhead{$\log\Phi^*(z=6)$\tablenotemark{a}} & 
  \colhead{$M_{1450}^*$} & 
  \colhead{$\alpha$} & 
  \colhead{$\beta$}
 }
 \startdata
 $-8.94^{+0.20}_{-0.24}$  &  $-27.21^{+0.27}_{-0.33}$  &  $-2.03^{+0.15}_{-0.14}$  &  -4.00  \\[4pt]
 $-8.10^{+0.30}_{-0.25}$  &  $-25.88^{+0.60}_{-0.49}$  &  -1.50  &  $-3.12^{+0.28}_{-0.41}$  \\[4pt]
 $-8.40^{+0.03}_{-0.03}$  &  -26.39  &  -1.80  &  -3.26 
 \enddata
 \tablecomments{Parameters without uncertainty ranges are fixed during the 
maximum likelihood fitting.}
\label{tab:mlefit}
 \tablenotetext{a}{$\log\Phi^*(z) = \log\Phi^*(z=6) + k(z-6)$, with $k=-0.47$.}
\end{deluxetable}

We now derive parametric fits to the observed data using maximum
likelihood estimation. 
The maximum likelihood
estimate for a luminosity function $\Phi(M,z)$ corresponds to the
minimum of the log likelihood function
\[
	S = -2 \sum\limits_i^N \ln[\Phi(M_i,z_i)] 
	      + 2\int\int\Phi(M,z)p(M,z)\frac{dV}{dz} dM dz ~,
\]
where the first sum is over the observed quasars, and the second
is over the full luminosity and redshift range of the sample and
provides the normalization (\citealt{Marshall+83}, see
\citealt{Fan+01PI} and \citealt{Kelly+08} for alternative
derivations of the likelihood function).
The term $p(M,z)$ is the probability
that a quasar with absolute magnitude $M$ (for our purposes, 
$M_{1450}$) and redshift $z$ is included in the survey; i.e.,
the selection function as derived in section~\ref{sec:completeness},
including all sources of incompleteness. 
Confidence intervals
are derived from the likelihood function using a $\chi^2$ distribution
in $\Delta S = S-S_{\rm min}$ \citep{Lampton+76}.

Over a wide
redshift range, the quasar luminosity function is found to be well
fit by a double power law \citep{BSP88},
\[
	\Phi(M,z) = \frac{\Phi^*(z)}
	     {10^{0.4(\alpha+1)(M-M^*)} + 10^{0.4(\beta+1)(M-M^*)}} ~,
\]
where $M^*$ is the characteristic luminosity at which the function
changes slope from steep at the bright end ($\beta$) to shallow
at the faint end ($\alpha$). The four QLF parameters may evolve with
redshift, possibly in an interdependent manner. Given the limited
redshift range of our survey, we will only account for the
steep decline in number density at high redshift using the
fit of \citet{Fan+01LF}: $\Phi^*(z) = \Phi^*(z=6)\times10^{k(z-6)}$, 
with $k=-0.47$.\footnote{We normalize $\Phi^*$ to $z=6$ for easier 
comparison to the higher redshift results.}

Even with a sample of nearly 200 quasars spanning $\Delta M \approx4$,
there are substantial degeneracies in fitting the data with this
parameterization. In particular, there are strong covariances between
the placement of the break luminosity and the indexes of the power-law slopes.
We thus perform several fits while fixing one or more parameters. 

Table~\ref{tab:mlefit} and Figure~\ref{fig:QLF_fits} show the results of 
several model fits with various choices for the fixed parameters.  
Uncertainties derived by varying a single parameter and calculating the 
likelihood for the best-fit solution for the other parameters are included. 
First, we fix the bright end slope to $\beta=-4.0$. This value was chosen as 
it is approximately the slope derived from a single power-law fit to the
brightest magnitude bins (\S\ref{sec:binnedqlf}). We choose to fix the bright
end slope as we find 
that our fits prefer high values for the break luminosity; interestingly, 
this implies that the \emph{bright end} slope is poorly constrained 
by our data (due to small numbers and limited luminosity range). 
This fit has a steep faint end slope and high break luminosity
($\alpha=-2.0$ and $M_{1450}^*=-27.2$). We adopt these values as our best fit.
Although the likelihoods can be improved by allowing even steeper values for 
$\beta$, this tends to drive the break luminosity near the limit of our data,
where both parameters have considerable freedom while fitting.

\begin{figure}
 \epsscale{1.15}
 \plotone{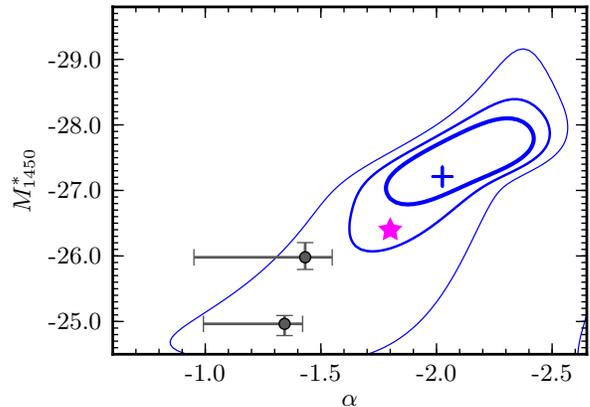}
 \caption{Likelihood contours for $\alpha$ and $M_{1450}^*$, with
 $\Phi^*$ and $\beta$ allowed to vary.
 Contours are drawn at the 68.3\%, 95.4\%, and 99.7\% confidence
 intervals, with decreasing line thickness. 
 The contours are relative to the peak likelihood for a
 model with $\beta=4.0$, which we have adopted as our best fit
 (indicated with a cross).
 The magenta star shows the best-fit
 values from \citet{Willott+10} for $\alpha=-1.8$. The two
 gray points with error bars represent the $z=2.2$ (lower) and $z=3.4$
 (upper) fits to the redshift-binned BOSS data from \citet{Ross+12qlf}.
 The data indicate an evolution towards higher break luminosities and
 steeper faint-end slopes with redshift.
 \label{fig:alpha_Mstar}
 }
\end{figure}

\begin{figure}
 \epsscale{1.15}
 \plotone{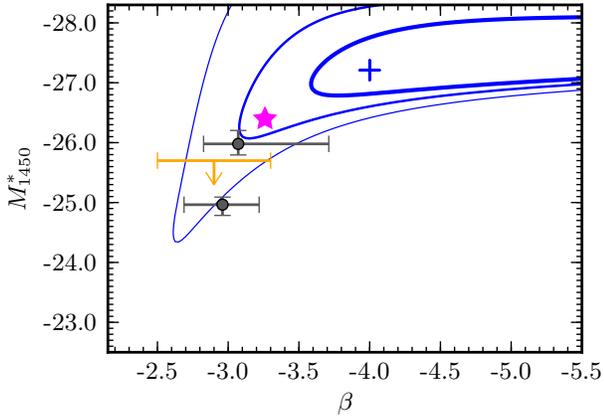}
 \caption{As in Fig.~\ref{fig:alpha_Mstar}, likelihood contours for $\beta$ 
 and $M_{1450}^*$ with comparison to values derived at other redshifts. 
 For comparison, we show the constraint from \citet{Richards+06LF} as the
 orange error bar ($M_{1450}^*$ is assigned an upper limit as it was not
 a free parameter in their fit). 
 The bright-end slope is consistent with little or no evolution from
 $z=2$ to $z=6$, although it is poorly constrained at all redshifts.
 \label{fig:beta_Mstar}
 }
\end{figure}

We further explore the parameter degeneracies by calculating the joint 
likelihood ranges for each of the power-law slopes and $M_{1450}^*$.
Figure~\ref{fig:alpha_Mstar} shows probability contours for $\alpha$ and 
$M_{1450}^*$, while allowing the other two parameters to vary.
Figure~\ref{fig:beta_Mstar} shows similar contours for $\beta$ and 
$M_{1450}^*$. Comparison of these two figures shows that $\beta$ is
indeed more poorly constrained than $\alpha$.

Surveys of faint $z\la3$ quasars typically find $\alpha=-1.5$
\citep{Hunt+04,Richards+05,Siana+08,Croom+09},
while observations at high redshift favor a steeper value of $\alpha=-1.7$
\citep{Glikman+10,Ikeda+11,Masters+12},
albeit with large uncertainties due to the difficulties of assembling large
samples of faint quasars at high redshift. We find that a steeper value for 
the faint end slope is favored by our data, although $\alpha=-1.5$ lies 
within our $3\sigma$ contour.

Figure~\ref{fig:beta_Mstar} shows that fits to our data similarly prefer
steep values of $\beta$. In fact, if we allow all of the parameters to be 
free, the maximum likelihood fitting tends to arbitrarily steep values for 
$\beta$. We consider fits with $\beta<-4$ to be effectively unconstrained, 
as this places the break luminosity at $M_{1450}^* \la -27.4$, where we have 
only a handful of quasars. We thus impose a ceiling on the likelihood based 
on a fit with $\beta=-4.0$, and calculate probability contours relative to 
this fit. Nonetheless, it is clear that $\beta$ is steep: $\beta<-3.1$
at 95\% confidence.
Further observations of bright quasars at $z\sim5$ are needed to better
constrain the bright-end slope at this redshift.

The strong constraints on the steepness of the bright end slope
show that a flattening of the bright end slope at high redshift,
as found by, e.g., \citet{Richards+06LF}, is not in agreement with
our data. This is likely due to the fact that \citet{Richards+06LF}
fit a single power law slope to the SDSS data under the assumption
that the break luminosity was well below their flux limit; we will
discuss this further in the following section.


\subsection{Evolution of the QLF}\label{sec:qlfevol}

\begin{figure}[!t]
 \epsscale{1.15}
 \plotone{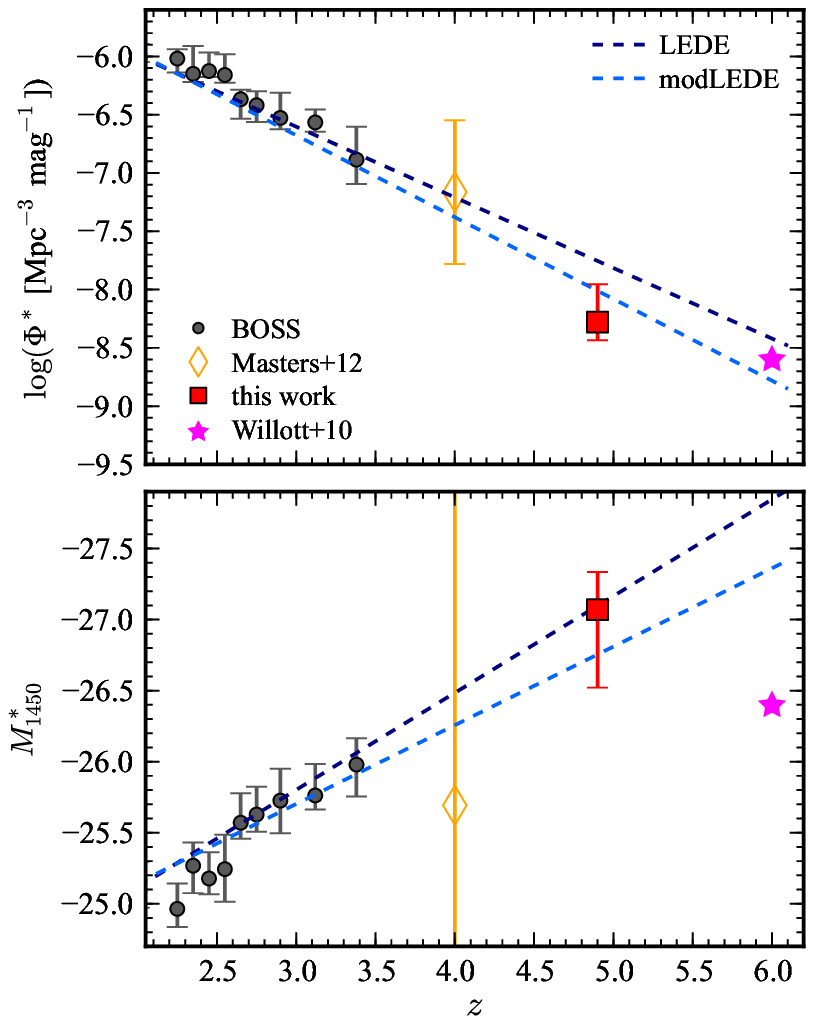}
 \caption{Evolution of the QLF normalization ($\Phi^*$) and break
 luminosity ($M_{1450}^*$) between $z\sim2$ and $z\sim5$. The
 points at $2.2<z<3.5$ come from the BOSS DR9 QLF \citep{Ross+12qlf},
 using a sample of $\sim6000$ variability-selected quasars from
 Stripe 82. The point at $z=4.9$ is from the best-fit model in 
 Table~\ref{tab:mlefit}. The point at $z=4$ is from \citet{Masters+12} and
 the one at $z=6$ is from \citet{Willott+10}, using their $\alpha=-1.8$ fit
 (uncertainties for the parameters were not reported for this fit).
 All points have been corrected to match our cosmology.
 The log-linear LEDE model fit to the BOSS data is shown as a dark blue 
 dashed line; the modified form of this model discussed in
 \S\ref{sec:spacedensity} is shown as a light blue dashed line.
 \label{fig:bosscompare}
 }
\end{figure}

\begin{figure}[!ht]
 \epsscale{1.1}
 \plotone{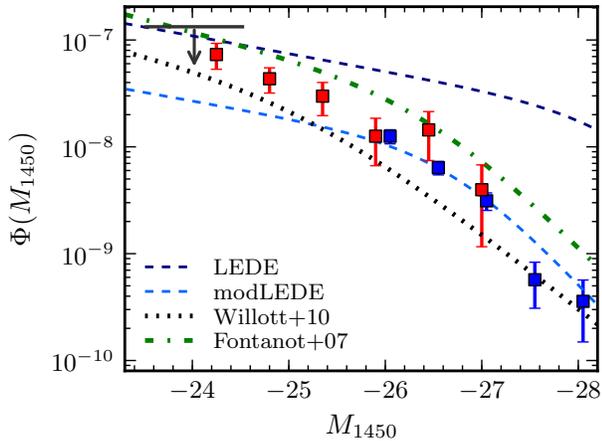}
 \caption{
 Evolutionary models for the QLF plotted against our binned data
 at $z=4.9$. The dark and light blue dashed lines are the BOSS LEDE
 and modified LEDE forms, respectively, as in Fig.~\ref{fig:bosscompare}.
 The LEDE model clearly disagrees with the data; the modified form
 agrees with the bright end reasonably well.
 The black dotted line represents the fit at $z=6$ from \citet{Willott+10} 
 shifted to $z=4.9$ using the \citet{Fan+01LF} redshift evolution;  
 the shape is roughly similar to our data but it clearly understimates the 
 normalization. The green dot-dashed line represents the QLF fit from
 \citet{Fontanot+07}. Also shown is the upper limit obtained by
 \citet{Ikeda+12} using COSMOS data (horizontal line with arrow).
 \label{fig:QLF_comparelede}
 }
\end{figure}

We provide some context for these fits by comparing to results
at lower and higher redshift. \citet{Ross+12qlf} recently presented a 
QLF measurement at $2.2<z<3.5$ from the BOSS DR9 quasar sample.
Over this redshift range, the QLF is well-fit by a 
Luminosity Evolution and Density Evolution (LEDE) model, where the
power law slopes have fixed values and the normalization and break
luminosity evolve in a log-linear fashion. Specifically,
\begin{eqnarray}
  \log[\Phi^{*}(z)]  & = & \log[\Phi^{*}(z=2.2)] + c_{1}(z-2.2)~,
  \label{eq:LEDE_linear_Phi} \\
            M_{i,2}^{*}(z) & = &M_{i,2}^{*}(z=2.2) + c_{2}(z-2.2)~.
  \label{eq:LEDE_linear_M}
\end{eqnarray} 
In equation~\ref{eq:LEDE_linear_M}, 
$M_{i,2} \equiv M_i(z=2) = M_{1450} - 1.486$ is the
absolute $i$-band magnitude at $z=2$ \citep{Richards+06LF}, 
corresponding to rest-frame $\sim2600$\AA\ and assuming
a spectral index of $\alpha_\nu=-0.5$ ($f_\nu \propto \nu^\alpha$).
The evolution in $M^*$ and $\Phi^*$ given by this model is shown in 
Figure~\ref{fig:bosscompare}, extrapolated to higher redshift
to compare with our data. 
While evolution in the power law slopes is not well constrained
by the data, our slopes are reasonably consistent with the values obtained
from the BOSS data, with some indication that the faint end
slope steepens toward higher redshift 
(Figs.~\ref{fig:alpha_Mstar}~and~\ref{fig:beta_Mstar}).
However, the values for $M^*$ and $\Phi^*$ do not agree with the
simple extrapolation of the LEDE model from lower redshift. This
can be clearly seen in Figure~\ref{fig:QLF_comparelede}, which shows
the LEDE prediction at $z=4.9$ significantly overestimates our measurements.
We will discuss a modified form to the LEDE model that provides
a better fit to the high redshift evolution in \S\ref{sec:spacedensity}.

At higher redshift, we compare to the $z\sim6$ QLF measurement
reported by \citet{Willott+10}. The last row of Table~\ref{tab:mlefit} shows 
the results of a fit where
all the parameters except $\Phi^*$ have been fixed to the best-fit
values from \citet{Willott+10} for a fixed faint end slope of 
$\alpha=-1.8$ (we use this value as it provides a better fit to our
data than their $\alpha=-1.5$ fit). 
Figures~\ref{fig:alpha_Mstar}~and~\ref{fig:beta_Mstar} show that the
$\alpha$,~$\beta$,~and~$M_{1450}^*$~from the \citet{Willott+10} QLF lie 
near the $\sim2\sigma$ contours from our constraints at $z=5$; 
however, the uncertainties on the \citet{Willott+10} values are not 
available and would likely eliminate any tension between the fitted values 
at $z=5$ and $z=6$ (see, e.g., their Figure~6 for the uncertainties on 
$\beta$ and $M_{1450}^*$ for a fit with $\alpha=-1.5$).
What is more clear is that the normalization is quite
different: it is a factor of $\sim1.6$ higher than predicted by
evolving the \citet{Willott+10} model to $z=4.9$ using $k=-0.47$, 
as they adopted for their fit (they obtained $\log\Phi^*(z=6)=-8.6$, in 
contrast to the value of $\log\Phi^*(z=6)=-8.4$ we obtain when fitting the 
shape of their QLF to our $z=5$ data).
Figure~\ref{fig:QLF_comparelede} illustrates the discrepancy in
the normalization from the evolved \citet{Willott+10} model; we 
will discuss this point further in \S~\ref{sec:spacedensity}.


The empirical QLF model of \citet[][hereafter HRH07]{HRH07} 
combines observations in 
the optical, X-ray, and mid-infrared bands to construct a bolometric 
QLF from $z=0$ to $z=5$. At high redshift the most constraining data 
in HRH07 comes from optical surveys, which have shown a 
flattening of the bright end slope at high redshift
\citep{SSG95,Fan+01LF,Richards+06LF}. In the HRH07 model,
the break luminosity increases with redshift until 
$z\sim2$ and then turns over, such that it is a factor of $\sim10$ 
lower luminosity at $z=5$ than at $z=2$.
At $z=5$, the HRH07 model predicts a quite 
low break luminosity ($M_{1450}^* \approx -22.6$)
and a shallow bright end slope ($\beta \approx -2.5$). 
Figure~\ref{fig:QLF_comparedata} shows that the HRH07 model
agrees with our data at $-27<M_{1450}<-26$, which is not surprising
since at $z\sim5$ their fit is mainly to the \citet{Richards+06LF}
data. The agreement at lower luminosities arises because the bright end
slope from HRH07 roughly agrees with our faint end slope.
At the bright end, we attribute the disagreement between our data and the 
HRH07 QLF to a steeper slope and a much brighter $M_{1450}^*$ than 
predicted by their model.
This demonstrates the substantial degeneracies in the QLF parameters,
as the HRH07 values for $\Phi^*$ and $M_{1450}^*$ at $z=5$ would not
even appear on Figure~\ref{fig:bosscompare}, even though the model
itself provides a good fit to our data at $M_{1450}\ga-27$. The key
difference between our work and HRH07 (and by extension,
\citealt{Richards+06LF}) is the increased survey area at the bright 
end, which is needed to extend {\em above} the break luminosity at
this redshift and make the break in the luminosity function more
evident.

Finally, we examine the $z=5$ QLF from \citet{Fontanot+07}. This
work combines bright quasars from SDSS DR3 with faint quasars from the
{\it Great Observatories Origins Deep Survey} \citep[GOODS;][]{goods} 
over the redshift range $3.5 < z < 5.2$. Figure~\ref{fig:QLF_comparelede} 
compares their pure density evolution (PDE) model (\#12 from their Table 3)
to our data. We find that the \citet{Fontanot+07} model overestimates our 
QLF measurements at all luminosities. \citet{Ikeda+12} similarly found some
tension between their constraint derived from observations of $z\sim5$
quasar candidates drawn from COSMOS \citep{cosmos}, and suggested this 
may be due to the completenesss correction applied by \citet{Fontanot+07}. 
We show the \citet{Ikeda+12} upper limit in Figure~\ref{fig:QLF_comparelede}.

In conclusion, we find evidence for a steepening
of the faint end slope, and no evidence in favor of an evolution
in the bright end slope, although obtaining strong constraints on the 
evolutionary forms of these parameters is difficult with existing data.
On the other hand, we do see evidence for strong evolution in the 
break luminosity, as it brightens from $M_{1450}^* \approx -25.4$ 
at $z=2.5$ to $M_{1450}^* \approx -27.2$ at $z=5$ 
(Figure~\ref{fig:bosscompare}).
This evolution has consequences for surveys where the faint
limit is near the break luminosity, as single power law fits
to such data would naturally find a flatter slope.
The problem is evident in Figure~\ref{fig:QLF_comparedata}, which
shows that a single power law can describe the SDSS DR3 data from 
\citet{Richards+06LF}, the full range of which is near the 
break luminosity.
The possibility that high redshift fits to the bright end of
the QLF may be biased by a higher break luminosity was put
forward by \citet{Assef+11} and \citet{SK12}. Based on our fits
at $z=5$ and the greater dynamic range of our survey, we
find this scenario to be plausible; i.e., the flattening of the
bright end slope at high redshift reported previously 
\citep{SSG95,Fan+01LF,Richards+06LF,HRH07} may be attributed instead
to rapid evolution in the break luminosity.

\subsection{Comparison to Theoretical Predictions}\label{sec:theorycompare}

\begin{figure}[!t]
 \epsscale{1.1}
 \plotone{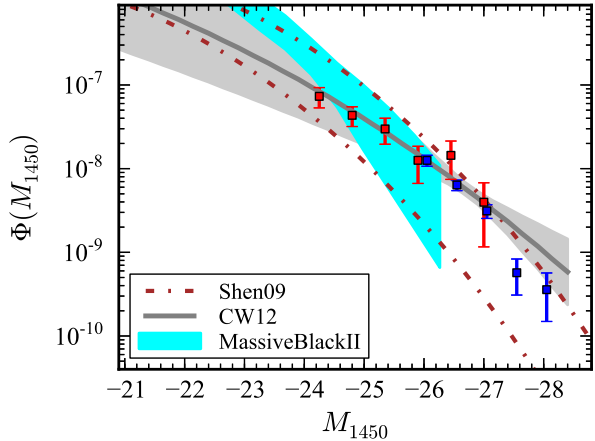}
 \caption{
 Comparison of observed QLF to various theoretical predictions.
 The two models of \citet{Shen09} are shown as dashed brown lines; the lower
 line is their fiducial model, and the upper line is their modified model
 as described in the text.
 The model of \citet{CW12} is represented by a solid gray line, with the
 shaded region indicating the $1\sigma$ uncertainty range.
 The range of the predicted QLF from the``MassiveBlackII'' 
 hydrodynamic simulations is represented by the cyan shaded region.
 \label{fig:QLF_comparemodels}
 }
\end{figure}

Figure~\ref{fig:QLF_comparemodels} compares our data 
to various theoretical models for the QLF at $z=5$. 
\citet{Shen09} provides a theoretical prediction for the evolution
of the QLF in a cosmological framework by relating the growth of
SMBHs to the hierarchical assembly of their host dark matter halos.
In this model, quasar activity is triggered by major mergers of 
halos. Their fiducial model reproduces the observed QLF at $0.5 < z < 4.5$, 
but underpredicts the observed QLF at higher 
redshifts. Indeed, their fiducial model lies below our data 
at $z=5$ (Figure~\ref{fig:QLF_comparemodels}).
\citet{Shen09} also has a variant of this fiducial model that 
includes a faster redshift 
evolution in the normalization of the scaling relation between peak quasar 
luminosity and host halo mass, a redshift evolution in the scatter of this 
relation and an increase in the upper host halo mass above which quasar 
triggering is cut off exponentially in order to better fit the QLF at 
higher redshifts \citep[see Sec. 4.4 of][for details]{Shen09}.
This alternative model provides a good fit to our data at the bright
end, but overpredicts the number counts at the faint end.

\citet{CW12} introduce a model in which quasars are tied to galaxies
in a straightforward manner through the $M_{\rm BH}~\mbox{--}~M_{\rm gal}$
relation, so that the evolution of the QLF is determined by the
evolution of this relation. Their model includes two free parameters
that are allowed to vary with redshift:
the quasar duty cycle and the normalization of the 
$M_{\rm BH}~\mbox{--}~M_{\rm gal}$ relation. \citet{CW12} constrain
these parameters from existing QLF measurements at $0.5 < z < 4.75$.
The high redshift constraints mainly come from the SDSS 
\citep{Richards+06LF} and have limited dynamic range, leading to
significant uncertainties in the $z\sim5$ prediction (represented
by the gray shaded region in Figure~\ref{fig:QLF_comparemodels},
see also Fig. 3 in \citealt{CW12}).
Overall, the \citet{CW12} model provides a good fit to the data.

\citet{Degraf+10} present a QLF prediction based on hydrodynamic
simulations that include radiative cooling, star formation, black
holes, and feedback processes. This work has since been updated
to a larger simulation volume, with 100~$h^{-1}$~Mpc on a side and
2x1792$^3$ particles (MassiveBlackII; DeGraf et al, in prep.). 
At each timestep, the QLF is calculated from the active black holes, and 
the final QLF prediction is derived by time-averaging the individual
measurements. This allows the bright end of the QLF to be estimated
by catching the brief episodes of peak luminosity among the rare,
massive black hole population. Figure~\ref{fig:QLF_comparemodels} shows
the prediction of this model in our redshift range, including
an estimate for cosmic variance derived by comparing two
simulation volumes (represented by the extent of the shaded region).
This model generally agrees with our data near and just below
the break luminosity. It appears to be somewhat steeper than the
data at faint luminosities, though fainter measurements of the
QLF from deeper optical data or at other wavelengths (e.g., X-rays)
are needed to better constrain the model.

We remind the reader that our QLF only accounts for unobscured,
Type I quasars. Our measurements are lower limits on the true density
of actively accreting black holes, assuming some fraction are in an
obscured (or even mildly extincted) growth phase. For example, comparison 
to X-ray surveys indicates that only $\sim25$\% of quasars are unobscured
at $z=4$ \citep{Masters+12}. Some inconsistency with theoretical models 
that do not distinguish between unobscured and obscured quasars is thus 
expected. Furthermore, a luminosity dependence for the obscured fraction
\citep[e.g.,][]{Ueda+03} could further bias comparisons of the data
with theoretical models.

\subsection{Spatial Density of Luminous Quasars}\label{sec:spacedensity}

A rapid decline in the comoving number density of quasars at
high redshift was observed three decades ago \citep{Osmer82}. 
Following \citet{Fan+01PI},  we quantify this evolution in terms 
of the spatial density of
quasars above a minimum luminosity within a redshift window. The
density is derived from the $1/V_a$ method, where
for each quasar the volume within which it would have been
observed within a survey is
\[
	V_a = \int_{\Delta z} p(M_{1450},z) \frac{dV}{dz} dz ~,
\]
where $p(M,z)$ is again the selection function for the survey.
From this equation, the total spatial density and its uncertainty are
estimated by
\[
	\rho = \sum_i\frac{1}{V_a^i} ~,~~ 
	\sigma(\rho)=
	  \left[\sum_i\left(\frac{1}{V_a^i}\right)^2\right]^{1/2} ~,
\]
where the sum is over all quasars more luminous than $M$.
This density is related to the luminosity function in that
\begin{equation}
    \rho(<M,z) = \int_{-\infty}^{M}\Phi(M,z) dM ~,
 \label{eqn:rhoM}
\end{equation}
where $\rho(<M,z)$ is the space density of quasars more luminous than $M$.
We choose to perform a sum over our data rather than an integration over
the QLF as the latter requires extrapolation and model fitting.
We calculate this quantity at $z\sim4$, $z\sim5$, and $z\sim6$, 
combining data from SDSS, our work, and multiple surveys at $z\sim6$. 
We choose a limit of $M_{1450}=-26$ as it corresponds to the lowest 
luminosity quasars in the SDSS sample at $z\sim4$.

At $z=4.25$ we use the uniform quasar
sample from the SDSS DR7 described in \S\ref{sec:dr7complete}. This sample 
includes 311 quasars with $4.1<z<4.4$, representing a
factor of four increase in number over the SDSS DR3 
results given in \citet{Richards+06LF}. We adopt the selection function
given in Table 1 of that work rather than recalculate it from our simulations,
as the SDSS quasar target selection  has a complicated dependence
on the extent of the stellar locus in color space \citep{Richards+02sel},
which is captured by the \citet{Richards+06LF} selection function.
At $z=4.9$ we use our combined sample from the SDSS DR7 and
from Stripe 82 and the selection functions presented in
section~\ref{sec:completeness}.
Finally, at $z=6$ we use the sample compiled by \citet{Willott+10},
consisting of quasars from the SDSS main 
\citep{Fan+01PI,Fan+03PII,Fan+04PIII,Fan+06PIV}, SDSS deep
\citep{Jiang+08,Jiang+09}, and CFHTQS \citep{Willott+10}. 
We derive selection
functions for the three $z\sim6$ quasar surveys using our quasar simulations,
extended to higher redshift.
The derived selection functions from our
models agree well with those shown in Figure 4 of \citet{Willott+10}.

\begin{figure}
 \epsscale{1.2}
 \plotone{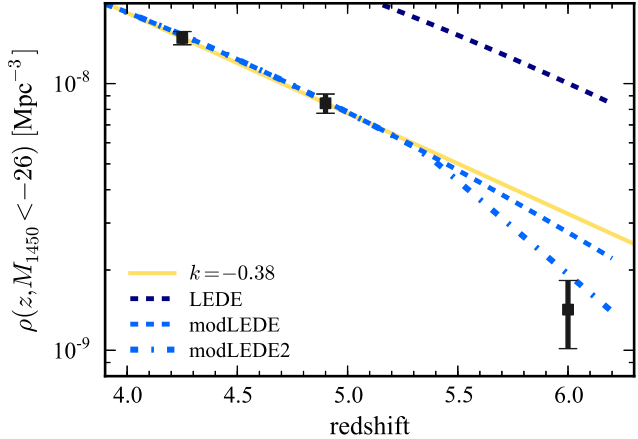}
 \caption{Evolution of the number density of luminous quasars at high
 redshift. The data are described in \S\ref{sec:spacedensity}.
 The yellow solid line shows the slope derived from a log-linear
 fit to the $z=4$ and $z=5$ points [$\log\rho(z) \propto -0.38z$],
 a value that is similar to the \citet{Fan+01LF} result based on
 39 bright quasars at $3.6<z<5$ from early Stripe 82 data.
 The $z=6$ point is well below this trend, departing from it
 at $4.5\sigma$.
 From $z=5$ to $z=6$ the number density declines
 as $\log\rho(z) = -0.7z$.
 The upper dark blue dashed line shows the extrapolated LEDE QLF model
 from the BOSS DR9, which clearly overpredicts the high redshift
 number densities.
 We modify the LEDE model to have a steeper evolution in $\log\Phi^*$
 and shallower evolution in $M_{1450}^*$ with redshift (light blue dashed line,
 see details in text), and further impose a maximum break luminosity of 
 $M_{1450}^*>-27$ (light blue dot-dashed line).
 \label{fig:zevol}
 }
\end{figure}

\begin{figure*}
 \epsscale{1.15}
 \plotone{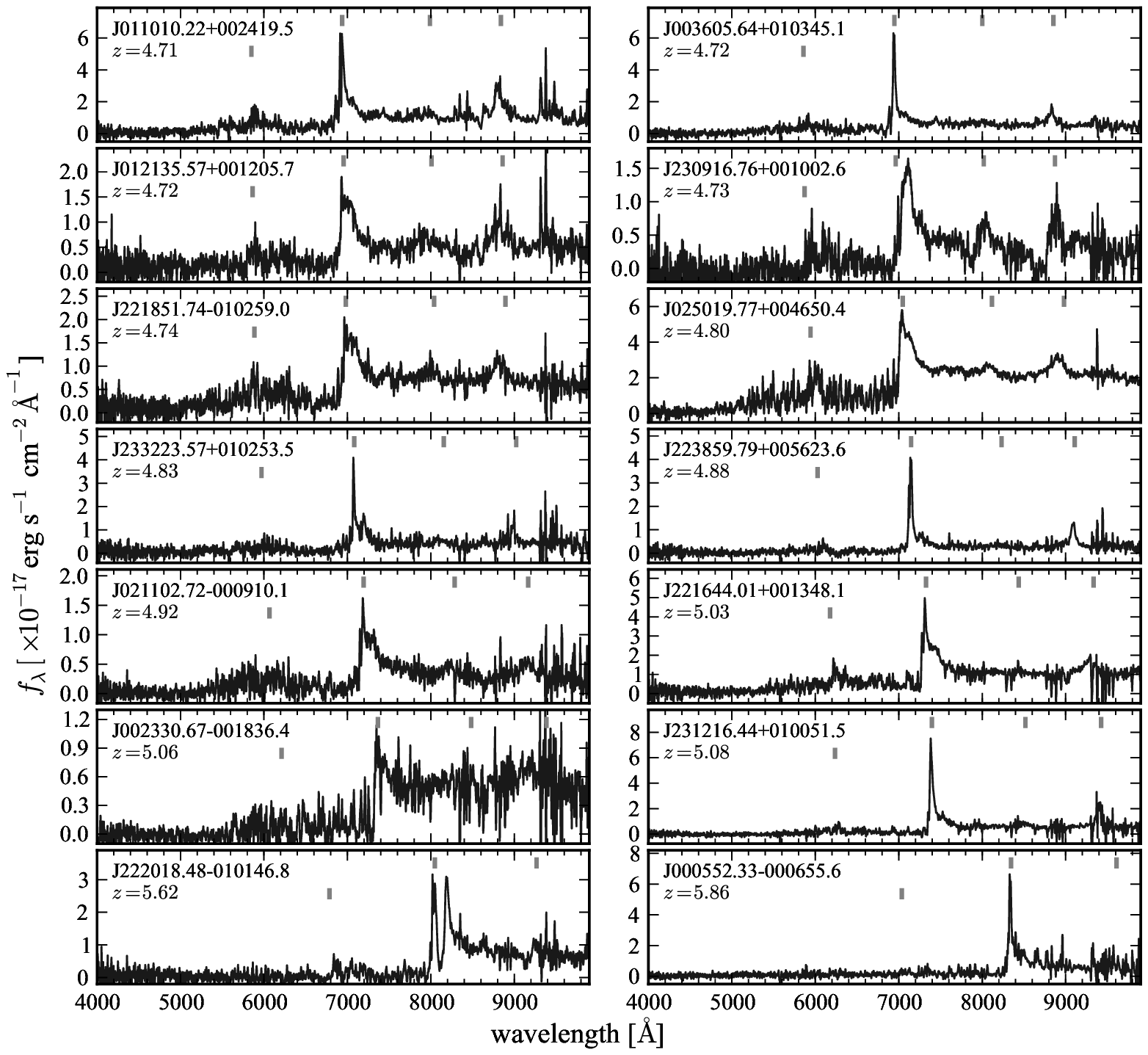}
 \caption{Spectra of $z>4.7$ Stripe 82 quasars from BOSS, 
 smoothed with a 5 pixel boxcar.
In addition to the coordinates and redshift, each plot
gives the plate-MJD-fiber designation for the BOSS observation.
The vertical gray lines mark the locations of typical emission
lines; in order, Ly$\beta$, Ly$\alpha$, \ion{Si}{4}, and \ion{C}{4}.
 \label{fig:bossspec}
 }
\end{figure*}

\begin{figure*}
 \epsscale{0.9}
 \plotone{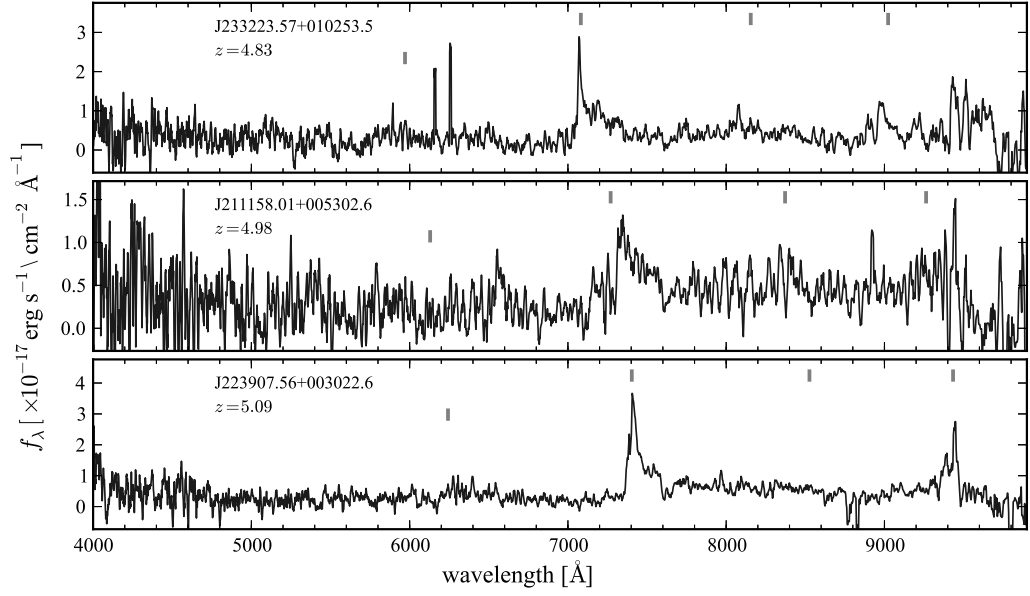}
 \caption{Spectra of Stripe 82 quasars observed with Magellan.
 The spectra have been rebinned to a resolution of $R\approx1200$
 and then smoothed with a 5 pixel ($\sim10$~\AA) boxcar.
 The first object is a repeat observation of an object with a 
 BOSS spectrum.
 Line markings are as in Fig.~\ref{fig:bossspec}.
 \label{fig:magespec}
 }
\end{figure*}

\begin{figure*}
 \epsscale{1.15}
 \plotone{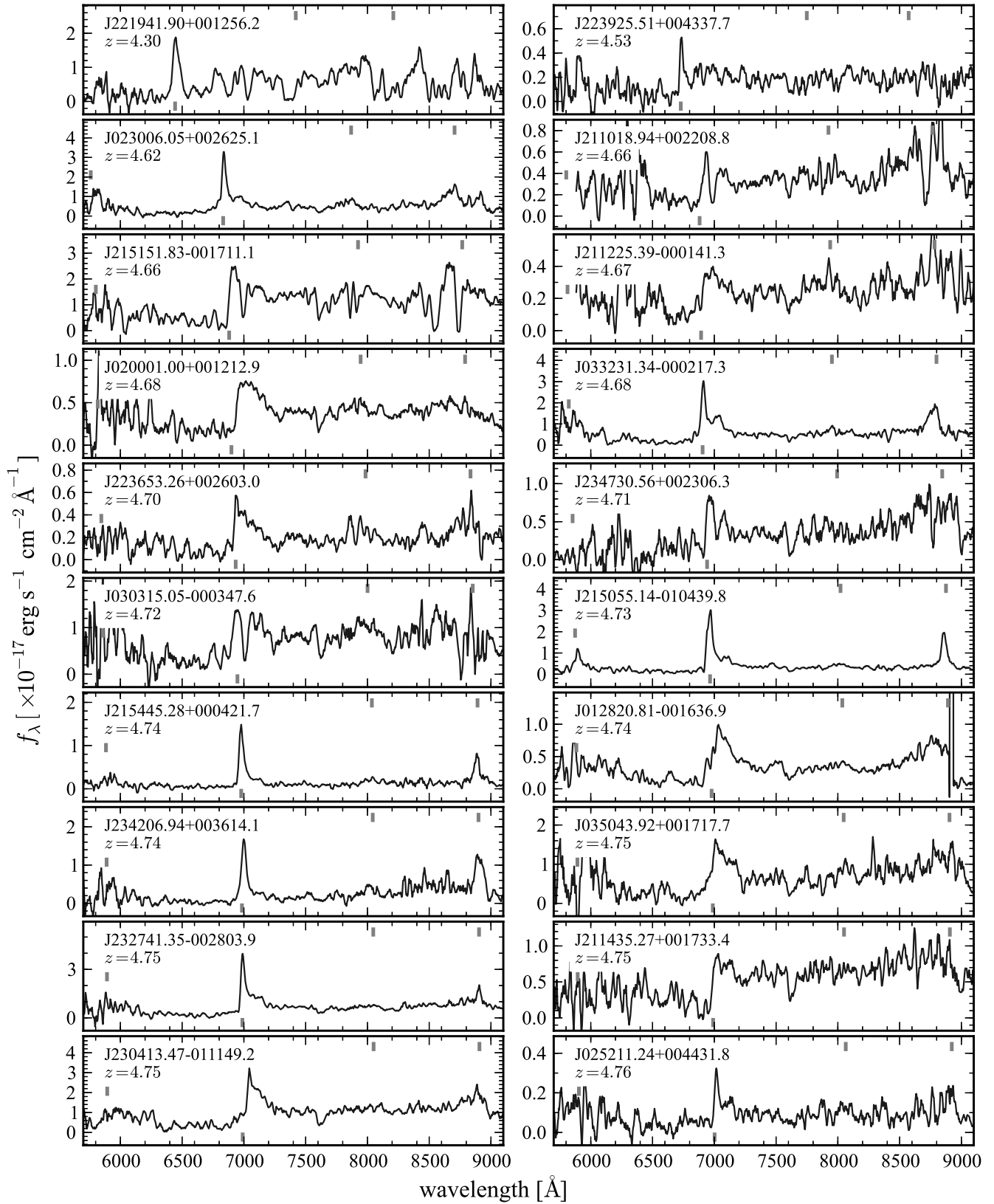}
 \caption{Spectra of Stripe 82 quasars observed with MMT Red Channel,
   smoothed with a 7 pixel boxcar. The MMT spectra are at low resolution
   ($R\sim400\mbox{--}600$) and from short, 10-15m. observations,
   thus they have relatively lower $S/N$. We present the spectra with with
   the nominal flux calibrations obtained from standard star observations;
   however, the absolute flux calibrations have considerable uncertainties.
    Line markings are as in Fig.~\ref{fig:bossspec}.
 \label{fig:mmtspec_1}
 }
\end{figure*}

Figure~\ref{fig:zevol} shows the evolution of the space density
of luminous quasars ($M_{1450}<-26$) from $z=4$ to $z=6$. 
Our measurement at $z=5$ 
shows a decline in the space density by a factor of 1.8 from $z=4.25$.
\citet{Fan+01LF} fit an exponential decline to the space density
at high redshifts, finding that $\rho(M_{1450}<-25.5,z) \sim 10^{kz}$ with
$k=-0.47$ at $z>3.6$,
about a factor of three per unit redshift. \citet{Brusa+09} reported
a similar value ($k=-0.43$) using X-ray-selected quasars at
$z\sim3.0\mbox{--}4.5$ ($\log L_X > 44.2$).
Our results from $z=4.25$ to $z=5$ are in good agreement
with these results; we measure a decline of $2.4 \pm 0.3$ per unit
redshift ($k=-0.38$).

On the other hand, the evolution to $z=6$ is more pronounced: 
a factor of $5.1 \pm 1.5$ per unit redshift ($k=-0.7$), or roughly 
twice the rate measured at lower redshift (Figure~\ref{fig:zevol}).
The departure of the $z=6$ measurement from the prediction based on the
slope fit to the $4 \la z \la 5$ data is significant at the $4.5\sigma$ level.
If we integrate deeper for the two high redshift bins, to a limit of
$M_{1450}<-24.5$, the decline is even stronger. We obtain
$\rho(M_{1450}<-24.5, z=4.9) = 4.2 \pm 0.7 \times 10^{-8}~\rm{Mpc}^{-3}$ and
$\rho(M_{1450}<-24.5, z=6.0) = 0.5 \pm 0.1 \times 10^{-8}~\rm{Mpc}^{-3}$,
corresponding to a factor of $6.5 \pm 1.7$ decline per unit redshift.

We check the consistency of our faint-end measurements by
comparing to recent work at X-ray wavelengths with the
{\sl Chandra}-COSMOS survey \citep{Civano+11}, which is sensitive to
moderate-luminosity 
($\log L_{(2-10~{\rm keV})} > 44.15~{\rm erg}~{\rm s}^{-1}$) AGN
at $z\sim5$. In order to compare the two samples, we must integrate our 
double power law model to $M_{1450}=-24.0$ 
\citep[adopting the $\alpha_{\rm ox}$ relation from][]{YER10}. 
\citet{Civano+11} identified three X-ray-selected Type I quasars in the redshift 
bin $4.2 < z < 5.4$, which translates to an observed space density of
$\Phi(4.2<z<5.4) =  (2.5 \pm 1.5) \times 10^{-7}~{\rm Mpc}^{-3}$.
Integrating our best-fit model for the QLF over that redshift bin, we obtain
$\Phi(4.2<z<5.4) =  0.8 \times 10^{-7}~{\rm Mpc}^{-3}$. Thus our QLF agrees 
with the \citet{Civano+11} results to within the $\sim1\sigma$ uncertainties, 
especially when considering that to make the comparison we must extrapolate 
our QLF in both luminosity and redshift. The X-ray data do not strongly
constrain the evolution to $z\sim6$ due to the small area of the 
{\sl Chandra}-COSMOS survey (from zero $z\sim6$ objects in
\citealt{Civano+11}, the decline from $z=5$ to $z=6$ is $>2$
per unit redshift).

We now relate the decline in the high-redshift quasar density to
the evolutionary model for the QLF derived from the BOSS DR9 
\citep{Ross+12qlf}.
In this LEDE model, $\log\Phi^*$ declines by 0.6 dex per unit
redshift, while $M^*$ brightens by 0.68 mag per unit redshift
($c_1=-0.6$ in equation~\ref{eq:LEDE_linear_Phi} and $c_2=-0.68$ in 
equation~\ref{eq:LEDE_linear_M}, respectively).
Integrating this model (equation~\ref{eqn:rhoM}) shows that it
significantly overpredicts the high redshift number densities
we have derived (Figure~\ref{fig:zevol}). This result suggests that
the steeper decline in the high-redshift number counts begins at
$z \la 4$, and is also consistent with Figure~\ref{fig:bosscompare},
which shows that the $\log\Phi^*$ predicted by the LEDE model
is higher than the data at $z \ga 5$, and is not well constrained
at $z\sim4$. We thus modify the LEDE model from BOSS by steepening
the slope of the $\log\Phi^*(z)$ evolution to $c_1=-0.7$, and softening 
the slope of the $M_{1450}^*$ evolution to $c_2=-0.55$. These 
modifications are within the $1\sigma$ uncertainties of the 
values derived from fitting BOSS data; Figure~\ref{fig:zevol} 
shows that they bring the QLF model in agreement with the 
integral constraints from the $z\sim4\mbox{--}5$ data, but
not the $z=6$ point. At $z=4.9$, this modified evolutionary model
predicts $M_{1450}^*=-26.8$ and $\log\Phi^*=-8.0$. This provides
a good match to the values we obtained when fixing the slopes 
to the BOSS values during the MLE fit, $M_{1450}^*=-26.6$ and 
$\log\Phi^*(z=4.9)=-7.9$ (Table~\ref{tab:mlefit}).

Finally, we consider further modifying the model for the evolution of
$M^*$. It is unlikely that $M^*$ continues to 
rise to very high luminosities; indeed, Figure~\ref{fig:bosscompare} 
shows that the high-redshift fits for $M^*$ are somewhat below the LEDE 
prediction. We thus impose a maximum break luminosity of 
$M_{1450}^*=-27$, which in the modified LEDE model is reached at 
$z\sim5.4$.
This choice causes a turnover in the high-redshift number densities that
matches the data at $z\sim6$ (Figure~\ref{fig:zevol}).
This model is somewhat arbitrary,
but qualitatively, a model in which $\Phi^*$ has a more rapid
downward evolution at high redshift and $M^*$ brightens until
$z\sim4\mbox{--}5$ and then levels off (or even turns over)
provides a reasonable description of the high redshift data.
A quasar with $M_{1450}^*=-27$ radiating at the Eddington
luminosity corresponds to a $\sim2\times10^9~M_\sun$ black hole;
it seems reasonable to suppose that while the break luminosity
evolves strongly at high redshift, it would not greatly exceed this
value.

\subsection{Contribution of $z\sim5$ quasars to the ionizing background}
\label{sec:ionizingbackground}

The number of ionizing photons required to maintain hydrogen ionization 
as a function of redshift can be found by balancing the ionizing 
photon emissivity with the density of hydrogen and the rate of 
recombinations.
The recombination rate depends on the clumping factor,
$C={\langle}n_H^2{\rangle} /{\langle}n_H{\rangle}^2$.
\citet{MHR99} present an equation for the ionizing photon density
$\dot{N}_{\rm ion}$ that adopts a high value for the clumping
factor ($C=30$).
More recent work suggests that the clumping factor is not
so large; \citet{Meiksin05} argues that $C \approx 5$ and 
\citet{MOF11} use cosmological simulations that include the 
effects of Lyman-limit systems and find $C \approx 2\mbox{--}3$
at $z=6$. Similarly, both \citet{Shull+12clump} and \citet{Finlator+12} 
prefer a lower clumping factor of $C \approx 3$ at $z\sim6$
in their reionization models.

We calculate the ionizing photon output for each quasar by assuming 
a broken power-law SED with an index of $\alpha_\nu=-1.7$ at ultraviolet
wavelengths \citep{Telfer+02}, a break at 1100\AA, and an index
of $\alpha_\nu=-0.5$ above the break\footnote{We also tested the far-UV
spectral slope from \citet{Shull+12uv} with $\alpha_{\nu}=-1.4$ and found
that it changes the results by only a few percent.}.
Rescaling the \citet{MHR99} equation and updating to our
cosmology, we estimate the number of photons required
to maintain full ionization at $z=5$ to be
$\dot{N}_{\rm ion} = 3.4\times10^{50} (C/5) ~\rm{Mpc}^{-3}~{\rm s}^{-1}$.
Integrating the best-fit QLF model (with $\beta=-4.0$) to $M_{1450}=-20$ gives
$\dot{N}_Q \sim 9.5 \times 10^{49}~\rm{Mpc}^{-3}~{\rm s}^{-1}$,
or $\approx 28$\% of the number required (for the flatter faint-end slopes,
the percentage lowers to $\sim20$\%).
For $C=2$, quasars provide $\sim70$\% of the photons required for
hydrogen ionization, suggesting they may have played some 
role in maintaining ionization at $z\sim5$.
However, the steeper decline of luminous quasars from $z\sim5$ 
to $z\sim6$ further reduces the likelihood that quasars were
an important source of ionizing photons during
the reionization epoch. 
Indeed, even when assuming a steep faint-end slope
($\alpha=-1.8$), \citet{Willott+10} find that $z\sim6$ quasars generate
$<10\%$ of the required ionizing photon background. This is also
in good agreement with constraints from deep X-ray surveys 
\citep{Barger+03,Fontanot+07} that limit the contribution from 
moderate luminosity AGN at $z>6$.

\section{Conclusions}\label{sec:conclusions}

This work builds on the legacy of SDSS Stripe 82 for high-redshift
quasar studies. \citet{Fan+04PIII} first used coadded photometry from
$\sim5$ epochs
of Stripe 82 imaging to discover a single $z\sim6$ quasar with
$z_{\rm AB} = 20.5$, fainter than the limit adopted for the single-epoch
imaging. 
Subsequently, \citet{Jiang+08,Jiang+09} created deep, coadded images
from the $50\mbox{--}60$ epochs available at the completion of 
SDSS I/II to discover 11 quasars at $z>5.8$ to a limit of
$z_{\rm AB}=22.8$. In this work, we have utilized the bluer SDSS 
bands to conduct a survey of $z\sim5$ quasars, a redshift where 
current constraints on the QLF are relatively weak. 

We define a sample of 92 candidates to a limit of $i_{\rm AB}=22.0$
using color selection criteria and with a well-defined selection
function. From this sample, 73 objects have spectroscopic
observations, and 71 are confirmed high redshift quasars. We then focus
on the redshift range $4.7 < z < 5.1$ where our completeness
is highest. Using a sample of 52 quasars from our work on Stripe 82
combined with 146 bright quasars from the SDSS DR7, we calculate 
the optical quasar luminosity function at $z=5$.

We have fit a double power law model to our observational data
at $z=5$, and reach the following conclusions:
\begin{itemize}
\item There is no clear evidence for evolution in the shape of the
      QLF from $z\sim2$ to $z\sim6$. We find a steep bright-end
      slope ($\beta \sim -4.0$, Fig.~\ref{fig:QLF_comparedata}) that 
      roughly agrees with measurements at both lower 
      \citep[e.g., BOSS $2.2<z<3.5$,][]{Ross+12qlf} and 
      higher \citep[$z=6$,][]{Willott+10,Jiang+09} redshifts. While
      the bright-end slope is not well determined due to small
      numbers, it is strongly constrained to be steep, with
      $\beta<3.1$ at 95\% confidence; thus we do not confirm previous
      findings that $\beta$ flattens at high redshift 
      \citep[e.g.,][]{Richards+06LF}.
      A relatively steep value for the faint end slope is favored 
      ($\alpha\approx-2$) in agreement with other high redshift results.
\item We see the break in the luminosity function in our $z\sim5$
      data, finding that $M_{1450}^*(z=4.9)=-27.2$ for
      $\beta=-4.0$ (Fig.~\ref{fig:beta_Mstar}~Table~\ref{tab:mlefit}).
      The strong covariances between the power law 
      slopes and the break luminosity lead to significant uncertainties
      in these quantities, but the best fit value for the break
      luminosity is signficantly higher than at $z=3.4$, where the BOSS
      results find $M_{1450}^*(z=3.4)=-26.0 \pm 0.2$.
\item The decline in the space density of luminous quasars at
      high redshift is greater than indicated by previous
      surveys. We find that while the 
      decline in the integrated space density of quasars 
      with $M_{1450}<-26$
      from $z\sim4$ to $z\sim5$ is about a factor
      of three per unit redshift, in agreement with previous
      results, the decline to $z\sim6$ is nearly a factor of
      two greater (Fig.~\ref{fig:zevol}). 
      This suggests a much more rapid dropoff
      in luminous quasar activity at the highest redshifts
      currently probed by observations.
\item By comparing to a simple LEDE model for the redshift
      evolution of BOSS quasars, we find that in addition to
      the steeper decline in number density, there is also an
      indication that the brightening of the break luminosity
      with redshift does not continue indefinitely. A toy model 
      in which the evolution of $\Phi^*$ is somewhat steeper than 
      from a simple extrapolation from the BOSS data, and in which
      the break luminosity evolves somewhat more slowly and peaks 
      at $M_{1450}^*=-27$, provides a good match to the high 
      redshift data.
\item Our model for the QLF at $z\sim5$ predicts that quasars
      contribute $\sim30\mbox{--}70$\% of the ionizing photons
      required to maintain hydrogen ionization at this redshift,
      with the uncertainty dominated by our lack of understanding
      of the clumping factor.
\end{itemize}

The strong evolution in the high redshift quasar number density
we have found has implications for quasar surveys at
even higher redshifts. Early forecasts for UKIDSS anticipated 
roughly 10 $z\sim7$ quasars from the survey \citep{WH02}, much greater
than the one found to date \citep{Mortlock+11} from over half the 
survey area. Using a redshift evolution of $\log\rho \propto -0.47z$
\citep{Fan+01LF}, extrapolation of our $z=5$ QLF predicts that there are
$\sim9$ quasars at $6.5<z<7.5$ in the UKIDSS DR8plus release to 
a limit of $Y({\rm AB})=20.2$ \citep[see Fig.~7 of][]{Mortlock+12}.
Using the same evolutionary factor, the $z=6$ QLF from \cite{Willott+10} 
predicts $\sim5$ quasars in this redshift range. Based on our results, 
these estimates should be revised downward to $\sim5$ and $\sim3$, 
respectively. The lower yield from UKIDSS is at least consistent with the
steeper evolution at $z>5$ we have found; further, the decline may continue 
to steepen at $z>6$, resulting in even smaller numbers of luminous quasars.

Significant progress has been made over the last decade in 
measuring the evolution of the high redshift quasar population.
Nonetheless, the observations are not yet strongly constraining
of models that make, for example, different predictions for the
evolution of the power law slopes. Ongoing high redshift quasar
surveys will improve these constraints. We are currently extending our
work to fainter luminosities at $z\sim5$ using deeper
optical imaging data, obtaining improved measurements of the faint
end slope and break luminosity. We will also use Stripe 82 data
to measure the QLF at $z\sim4$ over a wide range of luminosities.
Finally, Pan-STARRS \citep{Kaiser+02} is obtaining shallow optical 
and near-infrared imaging, including
the $y$-band, over an area of sky over that is more than twice that 
of the SDSS. To date, one $z\sim6$ quasar has been discovered from 
this survey \citep{Morganson+12}; future work should reduce the 
uncertainty on the bright end of the QLF at $z\sim5\mbox{--}6$ and better 
constrain the strong evolution we have measured from existing data.

\section{Acknowledgements}

The authors thank the staffs of the MMT and Magellan 
telescopes, particularly the recently retired John McAfee, for 
enabling many of the observations presented here.
IDM, LJ and XF acknowledge support from a David and Lucile Packard Fellowship, and NSF Grants AST 08-06861 and AST 11-07682.
L.J. acknowledges support from NASA through Hubble Fellowship grant HST-HF-51291.01 awarded by the STScI.

Funding for SDSS-III has been provided by the Alfred P. Sloan Foundation, the Participating Institutions, the National Science Foundation, and the U.S. Department of Energy Office of Science. The SDSS-III web site is http://www.sdss3.org/.

SDSS-III is managed by the Astrophysical Research Consortium for the Participating Institutions of the SDSS-III Collaboration including the University of Arizona, the Brazilian Participation Group, Brookhaven National Laboratory, University of Cambridge, Carnegie Mellon University, University of Florida, the French Participation Group, the German Participation Group, Harvard University, the Instituto de Astrofisica de Canarias, the Michigan State/Notre Dame/JINA Participation Group, Johns Hopkins University, Lawrence Berkeley National Laboratory, Max Planck Institute for Astrophysics, Max Planck Institute for Extraterrestrial Physics, New Mexico State University, New York University, Ohio State University, Pennsylvania State University, University of Portsmouth, Princeton University, the Spanish Participation Group, University of Tokyo, University of Utah, Vanderbilt University, University of Virginia, University of Washington, and Yale University. 

{\it Facilities:} 
 \facility{MMT (Red Channel spectrograph, SWIRC)}, 
 \facility{Magellan:Clay (MAGE)},
 \facility{SDSS}

\bibliographystyle{hapj}
\bibliography{z5stripe82}

\clearpage

\begin{figure*}
 \figurenum{22b}
 \epsscale{1.15}
 \plotone{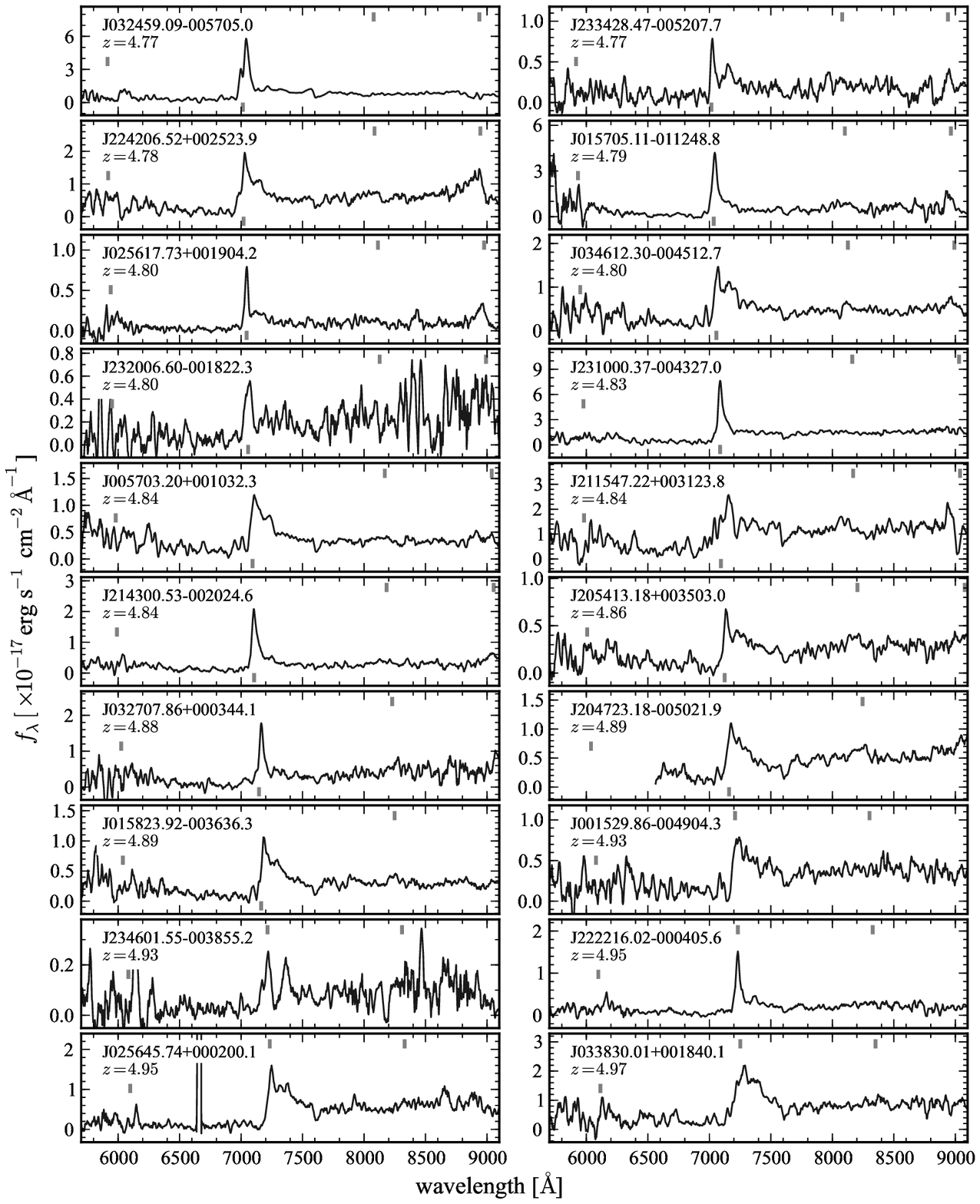}
 \caption{Spectra of Stripe 82 quasars observed with MMT Red Channel,
    continued.
 \label{fig:mmtspec_2}
 }
\end{figure*}
\begin{figure*}
 \figurenum{22c}
 \epsscale{1.15}
 \plotone{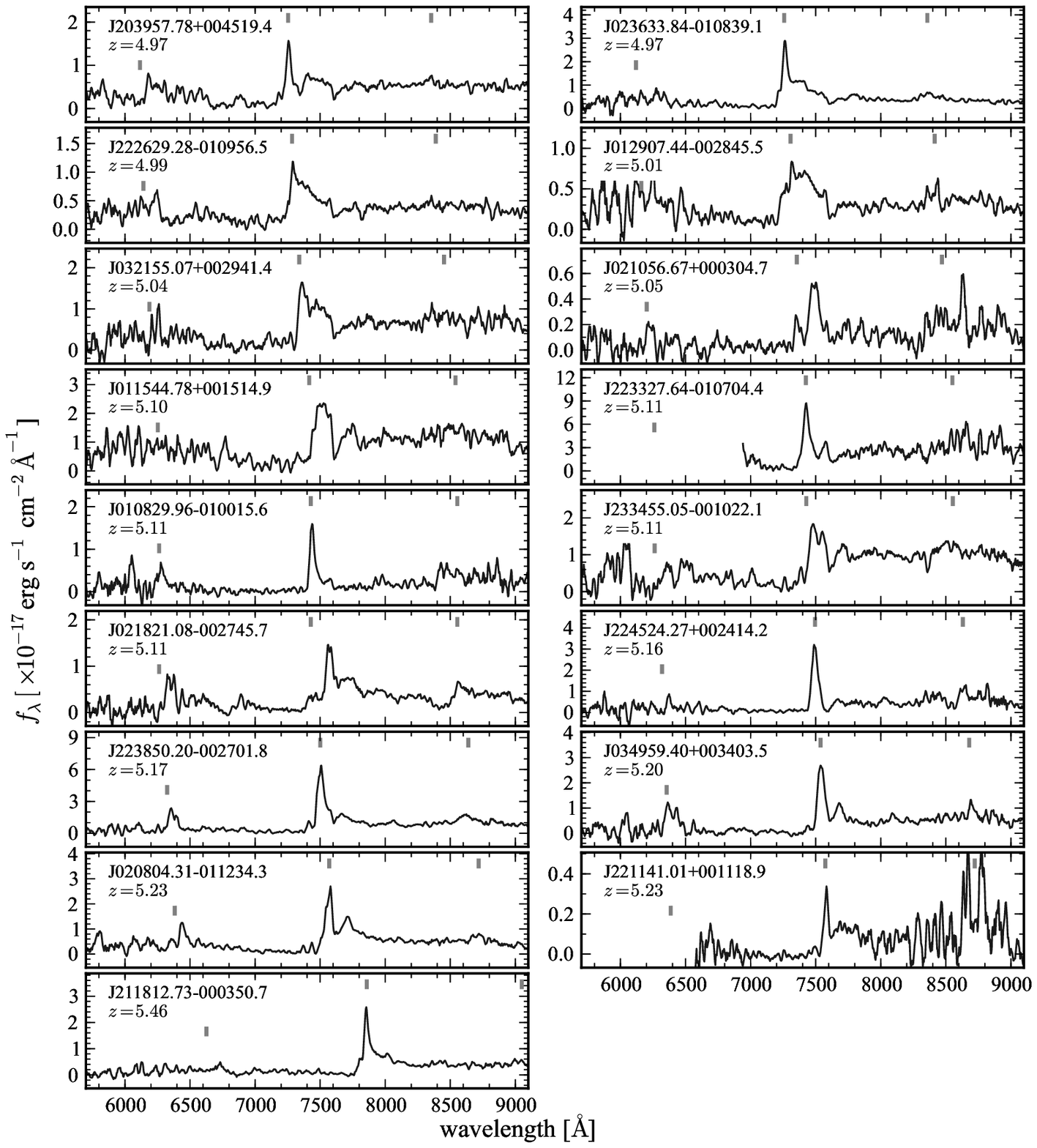}
 \caption{Spectra of Stripe 82 quasars observed with MMT Red Channel,
    continued.
 \label{fig:mmtspec_3}
 }
\end{figure*}

\clearpage
\LongTables
\begin{deluxetable*}{rrrrrrrrrrl}
 \centering
 \tablecaption{Stripe 82 high redshift quasar catalog}
 \tablewidth{0pt}
 \tablehead{
  \colhead{RA (J2000)} &
  \colhead{Dec (J2000)} &
  \colhead{$g_{\rm AB}$} &
  \colhead{$r_{\rm AB}$} &
  \colhead{$i_{\rm AB}$} &
  \colhead{$z_{\rm AB}$} &
  \colhead{$J_{\rm AB}$} &
  \colhead{$M_{\rm 1450}$} &
  \colhead{$z$} &
  \colhead{$p$\tablenotemark{a}} &
  \colhead{Notes\tablenotemark{b}} 
 }
 \startdata
20:39:57.78 & +00:45:19.4 &   &  $23.37 \pm 0.08$ &  $21.80 \pm 0.03$ &  $21.63 \pm 0.08$ &  \nodata &  -24.42 &  4.97 &  0.46 & 20120528 \\
20:47:23.18 & -00:50:21.9 &   &  $22.67 \pm 0.04$ &  $21.34 \pm 0.02$ &  $21.08 \pm 0.05$ &  \nodata &  -24.79 &  4.89 &  -1 & 20110624 \\
20:54:13.18 & +00:35:03.0 &   &  $22.83 \pm 0.05$ &  $21.59 \pm 0.02$ &  $21.46 \pm 0.07$ &  \nodata &  -24.50 &  4.86 &  0.74 & 20120528 \\
21:10:18.94 & +00:22:08.8 &  $24.99 \pm 0.26$ &  $22.79 \pm 0.05$ &  $21.32 \pm 0.02$ &  $21.14 \pm 0.06$ &  \nodata &  -24.63 &  4.66 &  -1 & 20110623 \\
21:11:58.01 & +00:53:02.6 &   &  $22.76 \pm 0.05$ &  $21.00 \pm 0.02$ &  $20.80 \pm 0.04$ &  $21.16 \pm 0.47$ &  -25.29 &  4.98 &  -1 & 20110611 \\
21:12:25.39 & -00:01:41.3 &  $24.93 \pm 0.23$ &  $22.42 \pm 0.03$ &  $20.98 \pm 0.01$ &  $20.81 \pm 0.04$ &  \nodata &  -24.99 &  4.67 &  -1 & 20110623 \\
21:14:35.27 & +00:17:33.4 &  $25.81 \pm 0.53$ &  $22.64 \pm 0.04$ &  $21.11 \pm 0.01$ &  $20.93 \pm 0.05$ &  $20.32 \pm 0.06$\tablenotemark{c} &  -24.89 &  4.75 &  0.84 & 20111002 \\
21:15:47.22 & +00:31:23.8 &  $24.97 \pm 0.27$ &  $22.40 \pm 0.03$ &  $20.83 \pm 0.01$ &  $20.68 \pm 0.04$ &  \nodata &  -25.27 &  4.84 &  0.77 & 20120528 \\
21:18:12.73 & -00:03:50.7 &   &  $23.72 \pm 0.10$ &  $21.60 \pm 0.02$ &  $21.18 \pm 0.05$ &  \nodata &  -25.64 &  5.46 &  -1 & 20120528 \\
21:30:08.93 & +00:26:09.8 &  $26.41 \pm 0.79$ &  $22.41 \pm 0.03$ &  $20.21 \pm 0.01$ &  $20.48 \pm 0.03$ &  \nodata &  -26.08 &  4.95 &  0.62 & DR7 \\
21:43:00.53 & -00:20:24.6 &  $26.18 \pm 0.68$ &  $23.58 \pm 0.09$ &  $21.94 \pm 0.03$ &  $21.93 \pm 0.11$ &   &  -24.11 &  4.84 &  0.28 & 20120527 \\
21:50:55.14 & -01:04:39.8 &   &  $23.15 \pm 0.07$ &  $21.52 \pm 0.03$ &  $21.55 \pm 0.08$ &   &  -24.45 &  4.73 &  0.79 & 20120527 \\
21:51:51.83 & -00:17:11.1 &  $24.84 \pm 0.27$ &  $22.44 \pm 0.04$ &  $20.75 \pm 0.01$ &  $20.55 \pm 0.03$ &  $20.13 \pm 0.05$\tablenotemark{c} &  -25.23 &  4.66 &  -1 & 20111002 \\
21:54:45.28 & +00:04:21.7 &   &  $23.57 \pm 0.09$ &  $21.91 \pm 0.03$ &  $21.71 \pm 0.11$ &   &  -24.05 &  4.74 &  0.41 & 20120528 \\
22:00:08.66 & +00:17:44.9 &  $24.11 \pm 0.10$ &  $20.60 \pm 0.01$ &  $19.05 \pm 0.00$ &  $19.19 \pm 0.01$ &  $19.27 \pm 0.07$ &  -27.11 &  4.82 &  0.88 & DR7 \\
22:11:41.01 & +00:11:18.9 &   &  $24.31 \pm 0.17$ &  $21.88 \pm 0.03$ &  $21.83 \pm 0.12$ &   &  -24.84 &  5.23 &  -1 & 20110624 \\
22:12:51.50 & -00:42:30.7 &  $24.96 \pm 0.25$ &  $21.64 \pm 0.02$ &  $19.88 \pm 0.00$ &  $20.03 \pm 0.02$ &  $20.55 \pm 0.25$ &  -26.43 &  4.95 &  0.61 & DR7 \\
22:16:44.01 & +00:13:48.1 &   &  $21.94 \pm 0.02$ &  $20.34 \pm 0.01$ &  $20.18 \pm 0.02$ &  $20.25 \pm 0.17$ &  -26.07 &  5.03 &  0.40 & 4200-55499-975 \\
22:18:51.74 & -01:02:59.0 &  $25.97 \pm 0.86$ &  $22.37 \pm 0.03$ &  $20.79 \pm 0.01$ &  $20.68 \pm 0.03$ &  $20.83 \pm 0.31$ &  -25.23 &  4.74 &  0.84 & 4201-55443-370 \\
22:19:41.90 & +00:12:56.2 &  $25.37 \pm 0.33$ &  $22.93 \pm 0.05$ &  $21.50 \pm 0.02$ &  $21.29 \pm 0.06$ &  $21.05 \pm 0.10$\tablenotemark{c} &  -24.37 &  4.30 &  -1 & 20111002 \\
22:20:18.48 & -01:01:46.8 &   &  $24.23 \pm 0.16$ &  $21.65 \pm 0.02$ &  $20.34 \pm 0.02$ &  $20.42 \pm 0.19$ &  -25.83 &  5.62 &  -1 & 4201-55443-246 \\
22:20:50.80 & +00:19:59.1 &  $23.83 \pm 0.08$ &  $21.51 \pm 0.01$ &  $20.08 \pm 0.00$ &  $20.04 \pm 0.02$ &  $19.74 \pm 0.10$ &  -25.95 &  4.72 &  0.90 & DR7 \\
22:22:16.02 & -00:04:05.6 &  $26.61 \pm 0.98$ &  $23.81 \pm 0.11$ &  $21.90 \pm 0.03$ &  $21.83 \pm 0.09$ &  $21.41 \pm 0.40$ &  -24.29 &  4.95 &  0.32 & 20120528 \\
22:25:09.19 & -00:14:06.8 &  $22.73 \pm 0.03$ &  $20.41 \pm 0.00$ &  $18.98 \pm 0.00$ &  $18.79 \pm 0.01$ &  $18.86 \pm 0.05$ &  -27.28 &  4.89 &  0.72 & DR7 \\
22:26:29.28 & -01:09:56.5 &   &  $23.27 \pm 0.07$ &  $21.69 \pm 0.02$ &  $21.50 \pm 0.07$ &  $20.76 \pm 0.28$ &  -24.57 &  4.99 &  0.55 & 20120527 \\
22:33:27.64 & -01:07:04.4 &   &  $23.62 \pm 0.09$ &  $21.52 \pm 0.02$ &  $21.49 \pm 0.07$ &  \nodata &  -24.97 &  5.11 &  -1 & 20110624 \\
22:36:53.26 & +00:26:03.0 &  $26.16 \pm 0.66$ &  $23.63 \pm 0.09$ &  $22.02 \pm 0.03$ &  $22.10 \pm 0.11$ &  $22.10 \pm 0.76$ &  -23.92 &  4.70 &  -1 & 20120528 \\
22:38:50.20 & -00:27:01.8 &  $26.06 \pm 0.60$ &  $23.11 \pm 0.06$ &  $21.49 \pm 0.02$ &  $21.36 \pm 0.06$ &  $21.62 \pm 0.17$\tablenotemark{c} &  -25.13 &  5.17 &  -1 & 20111002 \\
22:38:59.79 & +00:56:23.6 &  $24.74 \pm 0.61$ &  $22.88 \pm 0.05$ &  $20.91 \pm 0.01$ &  $20.96 \pm 0.04$ &  $21.43 \pm 0.51$ &  -25.23 &  4.88 &  -1 & 4204-55470-710 \\
22:39:07.56 & +00:30:22.6 &   &  $22.97 \pm 0.05$ &  $21.24 \pm 0.02$ &  $21.19 \pm 0.05$ &  $21.04 \pm 0.34$ &  -25.23 &  5.09 &  0.33 & 20110613 \\
22:39:25.51 & +00:43:37.7 &  $26.10 \pm 0.65$ &  $23.23 \pm 0.06$ &  $21.88 \pm 0.03$ &  $21.81 \pm 0.08$ &   &  -24.11 &  4.53 &  -1 & 20120528 \\
22:42:06.52 & +00:25:23.9 &  $25.41 \pm 0.33$ &  $23.39 \pm 0.07$ &  $21.85 \pm 0.03$ &  $21.77 \pm 0.10$ &  $21.73 \pm 0.82$ &  -24.14 &  4.78 &  0.52 & 20111002 \\
22:45:24.27 & +00:24:14.2 &  $26.46 \pm 0.92$ &  $23.75 \pm 0.11$ &  $21.72 \pm 0.02$ &  $21.86 \pm 0.11$ &  $21.30 \pm 0.12$\tablenotemark{c} &  -24.88 &  5.16 &  -1 & 20111002 \\
23:04:13.47 & -01:11:49.2 &  $25.29 \pm 0.37$ &  $22.78 \pm 0.04$ &  $21.24 \pm 0.02$ &  $21.14 \pm 0.05$ &  \nodata &  -24.75 &  4.75 &  0.84 & 20111002 \\
23:09:16.76 & +00:10:02.6 &  $26.31 \pm 0.68$ &  $22.62 \pm 0.03$ &  $20.91 \pm 0.01$ &  $20.99 \pm 0.05$ &  $21.16 \pm 0.27$ &  -25.08 &  4.73 &  0.83 & 4208-55476-896 \\
23:10:00.37 & -00:43:27.0 &  $25.48 \pm 0.33$ &  $22.62 \pm 0.03$ &  $20.88 \pm 0.01$ &  $20.87 \pm 0.04$ &  \nodata &  -25.21 &  4.83 &  0.77 & 20111002 \\
23:12:16.44 & +01:00:51.5 &  $25.60 \pm 0.44$ &  $22.71 \pm 0.04$ &  $20.87 \pm 0.01$ &  $20.75 \pm 0.03$ &   &  -25.61 &  5.08 &  -1 & 4209-55478-714 \\
23:20:06.60 & -00:18:22.3 &  $26.08 \pm 0.54$ &  $23.48 \pm 0.07$ &  $21.58 \pm 0.02$ &  $21.73 \pm 0.08$ &  $21.25 \pm 0.39$ &  -24.44 &  4.80 &  0.77 & 20111002 \\
23:27:41.35 & -00:28:03.9 &   &  $23.64 \pm 0.08$ &  $21.80 \pm 0.02$ &  $21.84 \pm 0.08$ &   &  -24.17 &  4.75 &  0.59 & 20111002 \\
23:32:23.57 & +01:02:53.5 &   &  $22.83 \pm 0.04$ &  $21.12 \pm 0.01$ &  $20.99 \pm 0.04$ &  $20.66 \pm 0.23$ &  -24.95 &  4.83 &  -1 & 4212-55447-680 \\
23:34:28.47 & -00:52:07.7 &   &  $23.80 \pm 0.10$ &  $21.99 \pm 0.03$ &  $22.18 \pm 0.11$ &   &  -23.99 &  4.77 &  0.25 & 20120527 \\
23:34:55.05 & -00:10:22.1 &   &  $23.56 \pm 0.07$ &  $21.82 \pm 0.02$ &  $21.33 \pm 0.05$ &  $20.85 \pm 0.38$ &  -24.65 &  5.11 &  -1 & 20111002 \\
23:42:06.94 & +00:36:14.1 &   &  $23.30 \pm 0.06$ &  $21.72 \pm 0.02$ &  $21.74 \pm 0.08$ &  $21.39 \pm 0.13$\tablenotemark{c} &  -24.25 &  4.74 &  0.69 & 20111002 \\
23:46:01.55 & -00:38:55.2 &   &  $23.64 \pm 0.09$ &  $21.69 \pm 0.02$ &  $21.56 \pm 0.07$ &   &  -24.49 &  4.93 &  0.62 & 20111001 \\
23:47:30.56 & +00:23:06.3 &  $25.88 \pm 0.41$ &  $22.72 \pm 0.04$ &  $21.09 \pm 0.01$ &  $20.75 \pm 0.04$ &  \nodata &  -24.89 &  4.71 &  0.81 & 20111002 \\
00:05:52.33 & -00:06:55.6 &   &  $25.42 \pm 0.42$ &  $23.05 \pm 0.07$ &  $20.47 \pm 0.02$ &  \nodata &  -24.42 &  5.86 &  -1 & 4216-55477-019 \\
00:07:49.16 & +00:41:19.5 &  $25.94 \pm 0.54$ &  $21.46 \pm 0.01$ &  $20.03 \pm 0.00$ &  $19.93 \pm 0.01$ &  $20.34 \pm 0.12$ &  -26.10 &  4.83 &  0.87 & DR7 \\
00:15:29.86 & -00:49:04.3 &  $25.57 \pm 0.36$ &  $22.57 \pm 0.03$ &  $21.01 \pm 0.01$ &  $20.69 \pm 0.03$ &  $20.56 \pm 0.19$ &  -25.20 &  4.93 &  0.68 & 20111001 \\
00:23:30.67 & -00:18:36.4 &   &  $23.10 \pm 0.05$ &  $21.31 \pm 0.01$ &  $20.82 \pm 0.03$ &  $21.50 \pm 0.52$ &  -25.10 &  5.06 &  0.44 & 4220-55447-474 \\
00:35:25.29 & +00:40:02.8 &  $23.73 \pm 0.07$ &  $21.38 \pm 0.01$ &  $19.66 \pm 0.00$ &  $19.71 \pm 0.01$ &  $19.83 \pm 0.09$ &  -26.42 &  4.76 &  0.94 & DR7 \\
00:36:05.64 & +01:03:45.1 &  $25.17 \pm 0.32$ &  $22.18 \pm 0.02$ &  $20.83 \pm 0.01$ &  $20.77 \pm 0.03$ &  $20.78 \pm 0.19$ &  -25.17 &  4.72 &  -1 & 4221-55443-864 \\
00:54:21.42 & -01:09:21.6 &   &  $21.28 \pm 0.01$ &  $19.92 \pm 0.00$ &  $19.54 \pm 0.01$ &  $19.51 \pm 0.09$ &  -26.63 &  5.09 &  -1 & DR7 \\
00:57:03.20 & +00:10:32.3 &   &  $22.98 \pm 0.05$ &  $21.45 \pm 0.02$ &  $21.31 \pm 0.07$ &  $21.12 \pm 0.28$ &  -24.61 &  4.84 &  0.78 & 20111001 \\
01:08:29.96 & -01:00:15.6 &   &  $23.67 \pm 0.10$ &  $21.85 \pm 0.03$ &  $21.69 \pm 0.07$ &  $21.35 \pm 0.52$ &  -24.63 &  5.11 &  -1 & 20111002 \\
01:10:10.22 & +00:24:19.5 &  $25.07 \pm 0.21$ &  $22.14 \pm 0.02$ &  $20.73 \pm 0.01$ &  $20.66 \pm 0.03$ &  $21.37 \pm 0.47$ &  -25.26 &  4.71 &  0.80 & 4227-55481-528 \\
01:15:44.78 & +00:15:14.9 &  $26.10 \pm 0.58$ &  $22.97 \pm 0.04$ &  $21.47 \pm 0.02$ &  $21.09 \pm 0.05$ &  $20.59 \pm 0.06$\tablenotemark{c} &  -25.00 &  5.10 &  -1 & 20111003 \\
01:21:35.57 & +00:12:05.7 &  $24.51 \pm 0.12$ &  $22.37 \pm 0.03$ &  $20.98 \pm 0.01$ &  $21.02 \pm 0.05$ &  $21.24 \pm 0.45$ &  -25.01 &  4.72 &  0.83 & 4228-55484-774 \\
01:28:20.81 & -00:16:36.9 &  $25.39 \pm 0.27$ &  $22.65 \pm 0.03$ &  $21.20 \pm 0.01$ &  $21.26 \pm 0.05$ &  $21.12 \pm 0.48$ &  -24.79 &  4.74 &  0.84 & 20111001 \\
01:29:07.44 & -00:28:45.5 &   &  $23.02 \pm 0.04$ &  $21.54 \pm 0.02$ &  $21.34 \pm 0.05$ &  $21.31 \pm 0.45$ &  -24.77 &  5.01 &  0.58 & 20111001 \\
01:57:05.11 & -01:12:48.8 &  $25.35 \pm 0.33$ &  $23.73 \pm 0.08$ &  $21.92 \pm 0.03$ &  $21.98 \pm 0.09$ &  $21.19 \pm 0.47$ &  -24.08 &  4.79 &  -1 & 20111003 \\
01:58:23.92 & -00:36:36.3 &   &  $23.30 \pm 0.05$ &  $21.62 \pm 0.02$ &  $21.39 \pm 0.05$ &   &  -24.50 &  4.89 &  0.70 & 20111001 \\
02:00:01.00 & +00:12:12.9 &   &  $23.15 \pm 0.04$ &  $21.64 \pm 0.02$ &  $21.40 \pm 0.06$ &  $21.21 \pm 0.11$\tablenotemark{c} &  -24.31 &  4.68 &  -1 & 20111001 \\
02:08:04.31 & -01:12:34.3 &   &  $23.12 \pm 0.05$ &  $21.46 \pm 0.02$ &  $21.21 \pm 0.04$ &   &  -25.28 &  5.23 &  -1 & 20111001 \\
02:10:43.16 & -00:18:18.4 &  $22.69 \pm 0.02$ &  $20.43 \pm 0.00$ &  $19.18 \pm 0.00$ &  $19.09 \pm 0.01$ &  $19.33 \pm 0.06$ &  -26.90 &  4.73 &  0.93 & DR7 \\
02:10:56.67 & +00:03:04.7 &   &  $23.97 \pm 0.10$ &  $21.60 \pm 0.02$ &  $21.41 \pm 0.07$ &  $21.17 \pm 0.10$\tablenotemark{c} &  -24.77 &  5.05 &  0.47 & 20111003 \\
02:11:02.72 & -00:09:10.1 &  $24.56 \pm 0.13$ &  $21.64 \pm 0.01$ &  $20.05 \pm 0.00$ &  $20.03 \pm 0.01$ &  $20.20 \pm 0.12$ &  -26.19 &  4.92 &  0.71 & 4236-55479-475 \\
02:18:21.08 & -00:27:45.7 &   &  $23.72 \pm 0.09$ &  $21.65 \pm 0.02$ &  $21.31 \pm 0.04$ &   &  -24.84 &  5.11 &  -1 & 20111001 \\
02:30:06.05 & +00:26:25.1 &  $25.39 \pm 0.25$ &  $22.94 \pm 0.04$ &  $21.69 \pm 0.02$ &  $21.46 \pm 0.05$ &  $21.65 \pm 0.59$ &  -24.26 &  4.62 &  -1 & 20111002 \\
02:36:33.84 & -01:08:39.1 &  $25.70 \pm 0.46$ &  $23.34 \pm 0.06$ &  $21.29 \pm 0.01$ &  $21.43 \pm 0.05$ &  $21.89 \pm 0.82$ &  -24.97 &  4.97 &  0.64 & 20111001 \\
02:50:19.77 & +00:46:50.4 &  $22.99 \pm 0.03$ &  $21.10 \pm 0.01$ &  $19.68 \pm 0.00$ &  $19.63 \pm 0.01$ &  $19.90 \pm 0.11$ &  -26.43 &  4.80 &  0.91 & 4241-55450-836 \\
02:52:11.24 & +00:44:31.8 &  $24.95 \pm 0.21$ &  $22.55 \pm 0.03$ &  $21.06 \pm 0.01$ &  $21.12 \pm 0.04$ &  $21.74 \pm 0.72$ &  -24.95 &  4.76 &  0.83 & 20111001 \\
02:56:17.73 & +00:19:04.2 &   &  $23.56 \pm 0.09$ &  $21.73 \pm 0.02$ &  $21.83 \pm 0.10$ &  $21.99 \pm 0.13$\tablenotemark{c} &  -24.27 &  4.80 &  0.66 & 20111001 \\
02:56:45.74 & +00:02:00.1 &   &  $23.02 \pm 0.05$ &  $21.58 \pm 0.02$ &  $21.54 \pm 0.07$ &  $21.15 \pm 0.36$ &  -24.63 &  4.95 &  0.67 & 20111002 \\
03:03:15.05 & -00:03:47.6 &  $24.80 \pm 0.20$ &  $23.07 \pm 0.05$ &  $21.44 \pm 0.02$ &  $21.40 \pm 0.05$ &  $21.12 \pm 0.38$ &  -24.53 &  4.72 &  -1 & 20111003 \\
03:10:36.97 & -00:14:57.0 &  $23.01 \pm 0.04$ &  $21.35 \pm 0.01$ &  $19.80 \pm 0.00$ &  $19.75 \pm 0.01$ &  $19.81 \pm 0.06$ &  -26.24 &  4.72 &  -1 & DR7 \\
03:21:55.07 & +00:29:41.4 &   &  $22.83 \pm 0.04$ &  $21.44 \pm 0.02$ &  $21.25 \pm 0.05$ &   &  -24.92 &  5.04 &  0.52 & 20111002 \\
03:24:59.09 & -00:57:05.0 &  $25.86 \pm 0.71$ &  $22.53 \pm 0.03$ &  $20.30 \pm 0.01$ &  $20.37 \pm 0.02$ &  $20.01 \pm 0.10$ &  -25.76 &  4.77 &  0.87 & 20111001 \\
03:27:07.86 & +00:03:44.1 &   &  $23.11 \pm 0.06$ &  $21.45 \pm 0.02$ &  $21.45 \pm 0.07$ &  $21.63 \pm 0.35$ &  -24.66 &  4.88 &  0.75 & 20111002 \\
03:32:31.34 & -00:02:17.3 &  $26.20 \pm 0.71$ &  $22.94 \pm 0.04$ &  $21.60 \pm 0.02$ &  $21.60 \pm 0.07$ &  $21.90 \pm 0.43$ &  -24.35 &  4.68 &  -1 & 20111002 \\
03:38:29.31 & +00:21:56.3 &  $25.76 \pm 0.46$ &  $21.39 \pm 0.01$ &  $19.84 \pm 0.00$ &  $19.60 \pm 0.01$ &  $19.83 \pm 0.08$ &  -26.61 &  5.03 &  0.35 & DR7 \\
03:38:30.01 & +00:18:40.1 &  $26.60 \pm 1.00$ &  $22.75 \pm 0.04$ &  $21.16 \pm 0.01$ &  $21.03 \pm 0.05$ &  $21.48 \pm 0.37$ &  -25.09 &  4.97 &  0.64 & 20111002 \\
03:46:12.30 & -00:45:12.7 &   &  $22.07 \pm 0.03$ &  $20.49 \pm 0.01$ &  $20.46 \pm 0.03$ &  $20.80 \pm 0.39$ &  -25.59 &  4.80 &  0.83 & 20111001 \\
03:49:59.40 & +00:34:03.5 &   &  $23.54 \pm 0.11$ &  $21.38 \pm 0.02$ &  $21.27 \pm 0.06$ &  $21.28 \pm 0.30$ &  -25.31 &  5.20 &  -1 & 20111002 \\
03:50:43.92 & +00:17:17.7 &  $25.27 \pm 0.47$ &  $22.56 \pm 0.05$ &  $21.28 \pm 0.02$ &  $21.43 \pm 0.10$ &   &  -24.72 &  4.75 &  0.84 & 20111003 
 \enddata
 \tablecomments{Magnitudes are on the AB system \citep{OkeGunn83} and 
have been corrected for extinction. Blank entries indicate objects for which
Sextractor did not report a magnitude in that band (due to negative fluxes);
objects that are not formally detected but have measurements from Sextractor
are included with their uncertainties. Ellipses indicate lack of coverage.}
\label{tab:quasars}
 \tablenotetext{a}{The selection probability $p(M,z)$ of spectroscopically
  confirming this object, including all sources of incompleteness 
  (see \S\ref{sec:s82complete}). A value of $-1$ indicates an object not 
  included in the uniform sample.}
 \tablenotetext{b}{Spectroscopic identifications from MMT/Magellan show
 the UT date of the observation; those from BOSS give the 
 plate-MJD-fiber of the observation.}
 \tablenotetext{c}{$J$ magnitude from SWIRC.}
\end{deluxetable*}
\clearpage

\begin{deluxetable*}{rrrrrrrrrrl}
 \centering
 \tablecaption{Stripe 82 high redshift quasar targets}
 \tablewidth{0pt}
 \tablehead{
  \colhead{RA (J2000)} &
  \colhead{Dec (J2000)} &
  \colhead{$g_{\rm AB}$} &
  \colhead{$r_{\rm AB}$} &
  \colhead{$i_{\rm AB}$} &
  \colhead{$z_{\rm AB}$} &
  \colhead{$J_{\rm AB}$} &
  \colhead{Notes}
 }
 \startdata
20:37:53.64 & +00:05:49.4 &   &  $22.81 \pm 0.05$ &  $20.84 \pm 0.01$ &  $20.33 \pm 0.03$ &  \nodata & \\
20:49:47.51 & +00:08:24.1 &  $25.82 \pm 0.59$ &  $23.38 \pm 0.09$ &  $21.91 \pm 0.03$ &  $21.64 \pm 0.11$ &  \nodata &  star \\
20:59:54.11 & -00:54:27.7 &  $25.90 \pm 0.76$ &  $23.11 \pm 0.06$ &  $21.54 \pm 0.02$ &  $21.24 \pm 0.05$ &  $21.22 \pm 0.15$\tablenotemark{a} & \\
21:00:41.31 & -00:52:03.4 &  $25.58 \pm 0.58$ &  $23.59 \pm 0.10$ &  $21.79 \pm 0.03$ &  $21.37 \pm 0.06$ &  $20.94 \pm 0.10$\tablenotemark{a} & \\
21:05:17.96 & -00:49:20.2 &   &  $23.23 \pm 0.07$ &  $21.49 \pm 0.02$ &  $21.18 \pm 0.06$ &  $20.86 \pm 0.31$ & \\
21:47:30.25 & +00:12:37.5 &  $25.32 \pm 0.38$ &  $23.65 \pm 0.11$ &  $21.92 \pm 0.03$ &  $21.50 \pm 0.09$ &  $20.82 \pm 0.38$ & \\
21:52:56.25 & -01:06:13.8 &  $25.01 \pm 0.46$ &  $23.00 \pm 0.07$ &  $21.56 \pm 0.03$ &  $21.32 \pm 0.07$ &  $20.66 \pm 0.08$\tablenotemark{a} & \\
00:03:32.25 & +00:37:49.9 &  $26.30 \pm 0.72$ &  $23.18 \pm 0.05$ &  $21.61 \pm 0.02$ &  $21.37 \pm 0.05$ &  $21.30 \pm 0.32$ & \\
00:05:31.71 & -00:34:43.4 &   &  $23.85 \pm 0.10$ &  $21.91 \pm 0.03$ &  $21.75 \pm 0.07$ &  $21.77 \pm 0.64$ & \\
00:18:22.10 & +00:25:23.1 &  $26.20 \pm 0.64$ &  $23.61 \pm 0.08$ &  $21.79 \pm 0.02$ &  $21.68 \pm 0.08$ &  \nodata & \\
00:23:53.63 & -00:43:56.9 &  $26.00 \pm 0.52$ &  $23.64 \pm 0.08$ &  $21.51 \pm 0.02$ &  $21.00 \pm 0.04$ &  $20.75 \pm 0.07$\tablenotemark{a} & \\
00:30:30.36 & -00:04:17.7 &  $25.78 \pm 0.42$ &  $23.41 \pm 0.06$ &  $21.78 \pm 0.02$ &  $21.67 \pm 0.06$ &  $21.69 \pm 0.76$ & \\
00:41:37.17 & -00:30:39.4 &   &  $22.94 \pm 0.04$ &  $20.92 \pm 0.01$ &  $20.45 \pm 0.02$ &  $20.00 \pm 0.15$ & \\
01:26:55.05 & -00:00:12.7 &   &  $24.37 \pm 0.16$ &  $21.87 \pm 0.02$ &  $21.78 \pm 0.10$ &  $21.09 \pm 0.40$ &  star \\
01:40:14.61 & -00:12:59.9 &  $25.42 \pm 0.30$ &  $23.30 \pm 0.05$ &  $22.00 \pm 0.02$ &  $21.81 \pm 0.07$ &  $21.47 \pm 0.60$ & \\
02:38:09.21 & -01:09:27.5 &  $28.23 \pm 4.72$ &  $23.44 \pm 0.06$ &  $21.94 \pm 0.03$ &  $21.83 \pm 0.08$ &  \nodata & \\
02:52:29.91 & -00:48:13.1 &   &  $23.74 \pm 0.09$ &  $21.97 \pm 0.03$ &  $21.77 \pm 0.08$ &  $21.14 \pm 0.41$ & \\
03:09:37.66 & +00:46:16.8 &  $26.77 \pm 1.16$ &  $23.32 \pm 0.06$ &  $21.91 \pm 0.03$ &  $21.74 \pm 0.07$ &  $21.38 \pm 0.37$ & \\
03:40:50.33 & +00:48:12.2 &  $25.99 \pm 0.61$ &  $23.23 \pm 0.07$ &  $21.88 \pm 0.03$ &  $21.73 \pm 0.07$ &  $21.06 \pm 0.23$ & \\
03:54:23.45 & +00:32:31.6 &  $24.00 \pm 0.44$ &  $22.92 \pm 0.15$ &  $21.28 \pm 0.04$ &  $21.05 \pm 0.08$ &  $20.33 \pm 0.18$ & \\
03:56:55.52 & -00:34:47.9 &  $24.82 \pm 0.65$ &  $22.48 \pm 0.08$ &  $21.21 \pm 0.03$ &  $21.09 \pm 0.07$ &  $22.01 \pm 1.16$ & 
 \enddata
 \tablecomments{Targets selected by our color criteria that either
lack spectroscopic observations or are confirmed non-quasars. Magnitudes
are given in the same format as Table~\ref{tab:quasars}.}
\label{tab:targets_noid}
 \tablenotetext{a}{$J$ magnitude from SWIRC.}
\end{deluxetable*}

\end{document}